\documentclass[a4paper,12pt]{article}
\usepackage{jheppub_modified_green}
\usepackage{epsfig}
\usepackage[makeroom]{cancel}
\usepackage{soul}
\usepackage{amsfonts}
\usepackage{amssymb,esint}
\usepackage{enumitem}
\usepackage{comment}
\usepackage{amsthm}
\usepackage{subfig}
\usepackage{bm}
\usepackage{hyperref}
\usepackage{mathrsfs}
\usepackage{cancel}
\usepackage{xcolor}
\usepackage{relsize}
\usepackage{float, slashed, graphicx, amssymb, amsmath}
\usepackage{bbm}
\usepackage{shuffle}
\usepackage{pgfplots}
\usepackage{tikz-feynman}
\pgfplotsset{compat=1.18}
\tikzfeynmanset{compat=1.1.0}
\tikzfeynmanset{graviton/.style={circle, draw=green!60, fill=green!5, very thick, minimum size=7mm}}
\tikzfeynmanset{codot/.style={/tikz/shape=circle,/tikz/fill=white,/tikz/minimum size=0.1cm,/tikz/inner sep=1.8pt}}
\tikzfeynmanset{myblob/.style={/tikz/shape=rectangle,/tikz/fill=red,/tikz/minimum size=0.2cm,/tikz/inner sep=1.8pt} }
\tikzfeynmanset{ghc/.style={/tikz/shape=circle,/tikz/fill=white,/tikz/minimum size=0.05cm,} }
\tikzfeynmanset{HV/.style=
{/tikz/shape=circle,
/tikz/fill={rgb:black,1;white,2},
 /tikz/minimum size=0.01cm,/tikz/inner sep=3.0pt
 } }
\tikzfeynmanset{GR/.style={/tikz/shape=ellipse,/tikz/fill={rgb:black,1;white,2},/tikz/minimum size=0.3cm,} }
\tikzfeynmanset{GR2/.style={/tikz/shape=ellipse,/tikz/fill={rgb:black,1;white,2},/tikz/minimum width = 1.2cm, 
    /tikz/minimum height = 3.4cm} }
\tikzset{box/.pic={\filldraw[fill=black]  (0,0) circle (2.5pt); \filldraw [fill=black] (0.5,0) circle (2.5pt); \draw [line width=5pt] (0,0) -- (0.5,0);}}
 \tikzfeynmanset{myblob2/.style=
{/tikz/shape=rectangle,
/tikz/fill=black,
 /tikz/minimum width=0.1cm,/tikz/inner sep=1.8pt
 } }
 \tikzfeynmanset{sb/.style=
{/tikz/shape=circle,
/tikz/fill=black,
 /tikz/minimum size=0.3cm,/tikz/inner sep=1.pt
 } }
 \tikzfeynmanset{myblob/.style=
{/tikz/shape=ellipse,
/tikz/fill=red,
 /tikz/minimum width=0.5cm,
 } }

\usetikzlibrary{arrows.meta} 
\usetikzlibrary{calc}
\usetikzlibrary{decorations.pathmorphing}
\usetikzlibrary{decorations.pathreplacing}
\usetikzlibrary{decorations.markings}
\tikzset{
   vector2/.style={decorate, decoration={snake, amplitude=1pt, segment length=6pt}, draw,double},
   vector/.style={decorate, decoration={snake, amplitude=1pt, segment length=6pt}, draw},
	provector/.style={decorate, decoration={snake,amplitude=2.5pt}, draw},
	antivector/.style={decorate, decoration={snake,amplitude=-2.5pt}, draw},
    fermion/.style={draw=black, postaction={decorate},
        decoration={markings,mark=at position .55 with {\arrow[draw=black]{>}}}},
    fermionbar/.style={draw=black, postaction={decorate},
        decoration={markings,mark=at position .55 with {\arrow[draw=black]{<}}}},
    fermionnoarrow/.style={draw=black},
    gluon/.style={decorate, draw=black,
        decoration={coil,amplitude=4pt, segment length=5pt}},
    scalar/.style={dashed,draw=black, postaction={decorate},
        decoration={markings,mark=at position .55 with {\arrow[draw=black]{>}}}},
    scalarbar/.style={dashed,draw=black, postaction={decorate},
        decoration={markings,mark=at position .55 with {\arrow[draw=black]{<}}}},
    scalarnoarrow/.style={dashed,draw=black},
    electron/.style={draw=black, postaction={decorate},
        decoration={markings,mark=at position .55 with {\arrow[draw=black]{>}}}},
	bigvector/.style={decorate, decoration={snake,amplitude=4pt}, draw},
}
\tikzset{cross/.style={cross out, draw, 
         minimum size=2*(#1-\pgflinewidth), 
         inner sep=0pt, outer sep=0pt}}

\tikzstyle{block} = [draw, rectangle, 
    minimum height=3em, minimum width=6em]

\usepackage[T1]{fontenc} 

\newcommand{\Tr}{\text{Tr}}

\def\sc#1{\overline{#1}}
\makeatletter
\newcommand \UPlus {\mathop {\operator@font \uplus }\limits }
\makeatother
\makeatletter
\newcommand \Bigcup {\mathop {\operator@font \bigcup }\limits }
\makeatother
\def\LabelNote#1{}
\def\Label#1{\label{#1}%
\smash{\hbox to\phipt{\raise1ex\hbox{\tiny[#1]}\hss}}}

\definecolor{bananayellow}{rgb}{1.0, 0.88, 0.21}
\definecolor{amber}{rgb}{1.0, 0.75, 0.0}

\newcommand{\cA}{\mathcal{A}}

\newcommand{\cF}{\mathcal{F}}
\newcommand{\cM}{\mathcal{M}}
\newcommand{\cC}{\mathcal{C}}

\newcommand{\cO}{\mathcal{O}}

\newcommand{\mb}{\bar m}

\newcommand{\ub}{\bar u}
\newcommand{\pb}{\bar p}

\def\nn{\nonumber}

\newcommand{\black}{\color{black}}

\def\bra#1{\langle #1|}
\def\ket#1{|#1 \rangle}

\def\braket#1{\langle #1 \rangle}

\def\spa#1.#2{\left\langle#1\,#2\right\rangle}
\def\spb#1.#2{\left[#1\,#2\right]}

\def\be{\begin{equation}}
\def\ee{\end{equation}}
\def\bea{\begin{eqnarray}}
\def\eea{\end{eqnarray}}  
\allowdisplaybreaks

\def\ep{\varepsilon}
\def\bep{\bm\varepsilon}

\usepackage{amssymb,amsmath}
\usepackage{mathtools} 
\usepackage{cancel} 
\usepackage{graphicx} 

\usepackage{hyperref}
\hypersetup{
	colorlinks=true,
	linktoc=page,
	citecolor=americanrose,
	linkcolor=cadmiumgreen,
	urlcolor=cadmiumgreen} 
\definecolor{americanrose}{rgb}{1.0, 0.01, 0.24}
\definecolor{cadmiumgreen}{rgb}{0.0, 0.42, 0.24}

\usepackage[colorinlistoftodos]{todonotes}

\newcommand{\Cdot}{{\cdot}} 
\makeatletter
\newcommand*{\bigcdot}{}
\DeclareRobustCommand*{\bigcdot}{%
  \mathbin{\mathpalette\bigcdot@{}}%
}
\newcommand*{\bigcdot@scalefactor}{.6}
\newcommand*{\bigcdot@widthfactor}{1.25}
\newcommand*{\bigcdot@}[2]{%
  \sbox0{$#1\vcenter{}$}
  \sbox2{$#1\cdot\m@th$}%
  \hbox to \bigcdot@widthfactor\wd2{%
    \hfil
    \raise\ht0\hbox{%
      \scalebox{\bigcdot@scalefactor}{%
        \lower\ht0\hbox{$#1\bullet\m@th$}%
      }%
    }%
    \hfil
  }%
}
\makeatother
\def\nn{\nonumber}

\title{Spinning binary dynamics in cubic effective field theories of gravity
}
\author{Andreas Brandhuber,}
\author{Graham R.~Brown,}
\author{Paolo Pichini,\\}
\author{\hspace{-0.2cm}Gabriele Travaglini}
\author{and Pablo  Vives Matasan}
\affiliation{Centre for Theoretical Physics, Department of Physics and Astronomy, \\
Queen Mary University of London, Mile End Road, London E1 4NS, United Kingdom}
\emailAdd{a.brandhuber@qmul.ac.uk}
\emailAdd{graham.brown@qmul.ac.uk}
\emailAdd{p.pichini@qmul.ac.uk}
\emailAdd{g.travaglini@qmul.ac.uk}
\emailAdd{p.vivesmatasan@qmul.ac.uk}
\begin{document}
\begin{flushright}
	QMUL-PH-24-09
\end{flushright}

\abstract{%
We study the binary dynamics of two Kerr black holes with arbitrary spin vectors in the presence of  parity-even and parity-odd cubic deformations of gravity. 
We first derive the 
tree-level
Compton amplitudes for a  Kerr black hole in cubic gravity, which we then use to compute the two-to-two amplitudes of the massive bodies to leading order in the deformation and the post-Minkowskian expansion. 
The required one-loop computations are performed
using the leading singularity approach as well as the heavy-mass effective field theory (HEFT) approach. These amplitudes  are then used to compute the leading-order momentum and spin kick in cubic gravity in the KMOC formalism.  
Our results are valid for generic masses and spin vectors, and include all the independent parity-even and  parity-odd cubic deformations of Einstein-Hilbert gravity. 
We also present spin-expanded expressions for the momentum and spin kicks, and the 
all-order in spin deflection angle in the case of aligned~spins.

}

\vspace{-2.6cm}

\maketitle

\flushbottom
 \tableofcontents
\newpage 

\section{Introduction}

 The detection of gravitational waves is entering a high-precision era, with  large numbers of gravitational-wave observations by the  LIGO-Virgo-KAGRA collaboration   
 and the prospect of increased  accuracy in forthcoming missions such as LISA. As a consequence there is a need for  ever more precise  theoretical predictions for the dynamics of binary black holes and  neutron stars. Available techniques to achieve this goal now include scattering amplitude methods,  developed over the years both in theoretical models and with a view to concrete applications to  high-precision measurements at the LHC.%
 \footnote{For a recent review see \cite{Travaglini:2022uwo}.} 
  By modelling  encounters of two celestial objects as a scattering  of two pointlike particles, modern amplitude methods can  be imported to study these processes in perturbation theory.  Crucial in  these derivations  is the observation \cite{Donoghue:1993eb} that only non-analytic terms in the amplitudes, arising from long-distance propagation of gravitons, are relevant to obtain the leading low-energy corrections, making unitarity-based methods perfectly suited for these calculations.   
 
Two natural avenues are available to explore. In the first, the focus is  solely on the Einstein-Hilbert (EH) action, and specifically on computing    perturbative corrections to various quantities of interest  to high orders in $GM/b$, where $G$ is Newton's constant, $b$~is the impact parameter and $M$ is the characteristic mass scale of the process.  Progress in this direction includes deriving Newton's potential  at 
 second Post-Minkowskian (PM) order 
 \cite{Neill:2013wsa,Bjerrum-Bohr:2013bxa, Bjerrum-Bohr:2014zsa, Bjerrum-Bohr:2016hpa}, the  potential at  3PM \cite{Bern:2019nnu,Bern:2019crd,Parra-Martinez:2020dzs,Cheung:2020gyp,Bjerrum-Bohr:2021din,DiVecchia:2021bdo,Brandhuber:2021eyq} and at    
 4PM \cite{Bern:2021dqo,Bern:2021yeh,Bern:2022jvn},   processes with radiation emission and waveforms 
\cite{Luna:2017dtq,Shen:2018ebu,Bautista:2019tdr,Herrmann:2021lqe,Herrmann:2021tct,Brandhuber:2023hhy, Herderschee:2023fxh, Elkhidir:2023dco,Georgoudis:2023lgf,Georgoudis:2023eke, Bohnenblust:2023qmy,Georgoudis:2024pdz,Bini:2024rsy,Brunello:2024ibk},     spin effects \cite{Vines:2017hyw,Guevara:2017csg,Arkani-Hamed:2017jhn,  Guevara:2018wpp,Chung:2018kqs,  Maybee:2019jus, Guevara:2019fsj,Arkani-Hamed:2019ymq,Damgaard:2019lfh,  Chung:2019duq, Aoude:2020onz,   Chung:2020rrz,Bern:2020buy, Guevara:2020xjx,  Bautista:2021wfy,Aoude:2021oqj,  Haddad:2021znf,Bautista:2021inx,Chen:2021kxt, Aoude:2022trd,Bern:2022kto, Alessio:2022kwv,Aoude:2022thd,Chen:2022clh,FebresCordero:2022jts,Menezes:2022tcs,Saketh:2022wap,Cangemi:2022abk,Damgaard:2022jem,       Comberiati:2022ldk,  Cangemi:2022bew,  Bautista:2022wjf, Bjerrum-Bohr:2023jau,Alessio:2023kgf,Bianchi:2023lrg,Bjerrum-Bohr:2023iey,  DeAngelis:2023lvf, Brandhuber:2023hhl, Aoude:2023dui, Cangemi:2023ysz,Gatica:2023iws, Bautista:2023sdf,Luna:2023uwd,  Cangemi:2023bpe,  Scheopner:2023rzp, Aoude:2023vdk, Haddad:2023ylx,   Gambino:2024uge}, 
and applications of the worldline formalism
\cite{Kalin:2020mvi, Kalin:2020fhe,Mogull:2020sak,Jakobsen:2021smu,Mougiakakos:2021ckm,Liu:2021zxr,Dlapa:2021npj,Jakobsen:2021lvp,Jakobsen:2021zvh,
Dlapa:2021vgp,Jakobsen:2022fcj,Riva:2022fru,Jakobsen:2022psy,Dlapa:2022lmu,Comberiati:2022cpm,Jakobsen:2022zsx,Jakobsen:2023ndj,Ben-Shahar:2023djm, Bhattacharyya:2024aeq}, which was recently used to obtain the first 5PM results 
\cite{Klemm:2024wtd, Driesse:2024xad}.  
Another direction entertains the possibility that general relativity is only the leading-order approximation of a yet to be discovered (more) complete theory. An efficient way to study this idea is through the framework of effective field theories (EFTs), initially developed in the context of gravity in  \cite{Donoghue:1993eb,Donoghue:1994dn}. 
In this language, higher-derivative operators are  added to the EH action, and their effect on  binary dynamics is  computed as a perturbative series in inverse powers of a scale, on which one can then  set bounds, see for example   \cite{Cardoso:2018ptl,Sennett:2019bpc,Silva:2022srr,Silva:2024ffz}.%
\footnote{In particular, in \cite{Silva:2022srr} a bound of $\ell_{\rm EFT}\leq 38.2$~km was found for the fundamental length scale of cubic theories, for the  choice $\beta_1 {=} \tilde{\beta}_1$, $\beta_2 {=} \tilde{\beta}_2{=}0$ for the couplings  in the effective action~\eqref{action}.}
%
One is then tasked to write down independent, gauge-invariant interactions of increasing mass dimension, which can  augment the EH action.

This approach was followed recently in \cite{Endlich:2017tqa}  for the case of dimension-eight operators. 
The first new couplings in fact arise at dimension four, that is   quadratic in the curvatures, but they  are known to  not affect scattering amplitudes \cite{Tseytlin:1986zz,Deser:1986xr,Tseytlin:1986ti} (see \cite{AccettulliHuber:2019jqo} for a recent derivation with modern amplitude methods). 
At dimensions six we encounter two independent parity-even  couplings together with  their two parity-odd cousins, which are cubic in the Riemann tensor. These cubic interactions    will be the focus of the present paper. They have been investigated already in a number of works, 
both using traditional relativity approaches%
\footnote{See also \cite{deRham:2019ctd,deRham:2020ejn,deRham:2021bll,CarrilloGonzalez:2022fwg,Melville:2024zjq} for related work in other modified theories of gravity.} \cite{Bueno:2016xff,Hennigar:2017ego,Silva:2022srr}, as well as 
amplitude techniques applied to the case of spinless binary systems \cite{Brandhuber:2019qpg, Emond:2019crr, AccettulliHuber:2020oou,AccettulliHuber:2020dal}, or systems where only one black hole is spinning and of much larger mass than the other \cite{Burger:2019wkq}.  
In particular, 
classical and quantum corrections to Newton's potential were computed at leading order in the perturbations \cite{Brandhuber:2019qpg, Emond:2019crr}, as well as the deflection angles for massless probes \cite{Brandhuber:2019qpg, AccettulliHuber:2020oou}. 
However,  in a generic encounter we expect both celestial bodies   to have non-zero spin vectors $a_1$ and $a_2$ and arbitrary  masses $m_1$ and $m_2$, and this is the situation we wish  to address in this work.

Cubic interactions are known to  give rise to time advance and hence causality violations already at tree level \cite{Camanho:2014apa}, but only at scales beyond the regime of validity of the effective field theory approximation \cite{deRham:2020zyh}, specifically for $b\lesssim \alpha^{1/4}$, where $\alpha$ is the typical coupling of a cubic interaction.%
\footnote{A detailed discussion of such violations in the amplitude framework for all parity-even cubic couplings can be found in \cite{AccettulliHuber:2020oou}.}
Therefore, cubic interactions  are the first nontrivial corrections to general relativity to  be included  in an  EFT  description  of gravitational phenomena. 
Specifically, the two independent parity-even ones can be chosen as    
\begin{align}
I_1 &\coloneq  {R^{\alpha \beta}}_{\mu \nu} {R^{\mu \nu}}_{\rho \sigma} {R^{\rho \sigma}}_{\alpha \beta}\ , \qquad 
I_2 \coloneq {R^{\mu \nu \alpha}}_\beta {R^{\beta \gamma}}_{\nu \sigma} {R^\sigma}_{\mu \gamma \alpha}\ , 
\end{align}
though, alternatively to $I_2$, the combination 
\begin{align}
    G_3 \coloneq I_1 - 2 I_2\   
\end{align}
is often considered. 
Therefore, in the following we will consider the 
gravitational effective action%
\footnote{In our normalisations, Newton's  constant is defined as   $G = \kappa^2 /(32\pi)$.}
\begin{equation}
\begin{split}
\label{action}
S = \int\!d^4x \sqrt{-g} \,  \bigg[& -\frac{2}{\kappa^2}R   \,  + \, \beta_1 I_1 + \beta_2 G_3 + \tilde{\beta}_1 \tilde{I}_1 + \tilde{\beta}_2 \tilde{G}_3\bigg]\, , 
\end{split}
\end{equation}
where $\tilde{I}_1$ and $\tilde{G}_3$ are defined as $I_1$ and $G_3$  except for replacing one Riemann tensor  by its dual $\tilde R^{\mu\nu\alpha\beta} {=} 
(1/2) \,\epsilon^{\mu\nu\rho\sigma} {R_{\rho\sigma}}^{\alpha\beta}$ in both terms. 
Notably, the interactions  $I_1$ and $G_3$ also appear in the low-energy effective action of bosonic string theory,  with the particular values 
\begin{equation}
\label{beta1beta2}
    \beta_1 = -\frac{2}{\kappa^2} \frac{\, \alpha^{\prime \, 2}}{48}\, , \qquad \beta_2 = -\frac{2}{\kappa^2} \frac{\, \alpha^{\prime \, 2}}{24}\, ,  \qquad\tilde{\beta}_{1,2}=0 ,  
\end{equation}
for the coefficients  $\beta_{1,2}$ and $\tilde{\beta}_{1,2}$. 
An  interesting feature of the $G_3$ operator is that it is  topological in six dimensions, and has vanishing four-dimensional  graviton amplitudes \cite{vanNieuwenhuizen:1976vb, AccettulliHuber:2020dal}. It does however lead to modifications of the classical  potential of a two-body system \cite{Brandhuber:2019qpg,Emond:2019crr}, although, unlike $I_1$, it does not cause deflection of  massless particles \cite{Brandhuber:2019qpg, AccettulliHuber:2020oou}. Furthermore, it can be reinterpreted as a tidal deformation after performing an appropriate field redefinition~\cite{AccettulliHuber:2020dal}.
We also note recent interest in parity violation, following  tentative indications in  the Cosmic Microwave Background and  the  
large-scale structure of galaxies, see e.g.~\cite{Minami:2020odp,Diego-Palazuelos:2022dsq,Eskilt:2022wav, Komatsu:2022nvu,Eskilt:2022cff,Philcox:2022hkh,Hou:2022wfj,Philcox:2023ffy}.

After describing what deformations we are interested in, we are left to discuss how to investigate their effect on observables. We will do so using  the KMOC formalism \cite{Kosower:2018adc}, further developed for spinning objects in \cite{Maybee:2019jus},  and soon after applied in \cite{Guevara:2019fsj} to  tree-level scattering of black holes in EH for arbitrary spin directions and to   all orders in the black-hole  multipole expansions.  
In particular we will compute, to leading order in the cubic couplings,  the momentum and spin kicks -- the variations in the momentum and spin of one of the two spinning bodies after the interaction has occurred.  
These quantities vanish at tree level, or first Post-Minkowskian order (1PM), and receive a non-vanishing classical contribution at one loop (2PM) which we derive. We obtain these quantities as functions of the masses, the spin vectors, the impact parameter and the relative velocity of the two objects, and  work in the post-Minkowskian expansion, with $G m_{1,2} {\ll} b$ and $|\vec{a}_{1,2}| {\lesssim} G m_{1,2}$,  and hence $|\vec{a}_{1,2}| {\ll} b$.

It turns out that at leading order in the deformations, the momentum and spin kicks can be extracted entirely from the Fourier transform to impact parameter space of the classical one-loop amplitude of four heavy spinning objects. The latter quantity is therefore what we have to compute. 
We will do so by exploiting recent progress on classical amplitudes of spinning particles \cite{Vaidya:2014kza,Arkani-Hamed:2017jhn,Johansson:2019dnu,Guevara:2018wpp,Chung:2018kqs,Vines:2018gqi,Siemonsen:2019dsu,Liu:2021zxr},  together with an application of   one-loop techniques  \cite{Cachazo:2017jef, Guevara:2017csg,Bautista:2023szu} based on triple cuts \cite{Forde:2007mi} or   the Heavy-mass Effective Field Theory (HEFT) approach developed in \cite{Brandhuber:2021kpo,Brandhuber:2021eyq,Brandhuber:2021bsf,Brandhuber:2022enp,Brandhuber:2023hhy}.

The first ingredients needed for our calculation are the spinning graviton Compton amplitudes in cubic gravity, which we  derive making use of the  energy-momentum tensor for a Kerr black hole  \cite{Vines:2017hyw,Chung:2018kqs,Guevara:2018wpp,Chung:2019yfs,Bern:2020buy,Bautista:2021wfy} and  the cubic vertices arising from $I_1$ and  $G_3$. The resulting Compton amplitudes are non-vanishing only  
for equal-helicity gravitons (for outgoing momenta), and are given in \eqref{I1amplitude} and \eqref{G3amplitude} for  $I_1$ and $G_3$, respectively. One notices the particularly simple form of  the latter amplitude, which is reflected in a correspondingly  simpler result for the one-loop amplitude \eqref{G3amp}, in comparison to the $I_1$ amplitude \eqref{I1amp}. The amplitudes from parity-odd interactions can be derived without a new calculation, and are closely related to those from the parity-even ones, see \eqref{parity-odd-triple-cut-result}. The results for the momentum and spin kicks, valid to all orders in the masses and spins of the heavy objects,   can be found in our   
\href{https://github.com/QMULAmplitudes/Cubic-Corrections-to-Spinning-Observables-from-Amplitudes}{{\it Cubic Corrections to Spinning Observables from Amplitudes} GitHub repository}. 
For the sake of presentation, we limit ourselves to quoting the explicit expressions of the  all-order in spin momentum kick,  the  linear in spin correction to the momentum and spin kicks, 
as well as the scattering angle for the case where the two black holes have aligned spins.  An interesting observation  in the case of aligned spins is that, for parity-odd interactions, the scattering is non-planar unlike for the parity-even deformations.

The rest of the paper is organised as follows.
In Section~\ref{sec: ClassicalComptonAmplitudes} we derive the classical gravitational Compton amplitudes for Kerr black holes using the corresponding classical energy-momentum tensor, together with the  cubic vertices induced by the new interactions. 
In Section~\ref{sec:1loopLS}, we use these gravitational Compton amplitudes to derive the classical one-loop  amplitudes for the two-to-two scattering of Kerr black holes with different masses and spins using triple cuts. In Section~\ref{sec:HEFT} we perform the same computation, this time using the HEFT formalism. 
The derivation for the parity-odd  interactions $\tilde{I}_1$ and $\tilde{G}_3$ is presented in  
Section~\ref{sec:parity-odd}. 
In Section~\ref{sec:Obs-from-KMOC}  we use the KMOC approach and the classical one-loop amplitudes derived earlier to compute the momentum and spin kicks to all orders in spin, and  later examine  the aligned spin case in detail.

Throughout this paper we will suppress the coefficients of the cubic interactions $\beta_1, \beta_2, \tilde{\beta}_1, \tilde{\beta_2}$, which  can always be reinstated at the end.

\section{Classical Compton amplitudes from cubic deformations}
\label{sec: ClassicalComptonAmplitudes}

We begin by quoting the expressions of the new three-point graviton vertices arising from $I_1$ and $I_2$, for gravitons of momenta $k_i$, $i=1,2,3$, where gravitons $1$ and $2$ are on shell, and $3$ is off shell with momentum $k_3=-k_1-k_2$.  For convenience we will write them as if they appear with coefficient $1$ in the action, i.e.~for $\beta_1=\beta_2=1$; at the end of the calculation these coupling constants can be set  to any value of interest, in particular to the choices \eqref{beta1beta2} arising from bosonic string theory.%
 \footnote{We also mention that $R^3$ graviton amplitudes were first considered in \cite{Broedel:2012rc}.} 
Throughout this section we focus on the parity-even interactions, and will discuss the Compton amplitudes arising from the parity-odd ones in Section~\ref{sec:parity-odd}. 

We assume that  the polarisation tensors have the factorised form $\varepsilon_{k_i}^{\mu\nu} = \varepsilon_{k_i}^\mu \varepsilon_{k_i}^\nu$ and obey the usual transversality conditions $\varepsilon_{k_i} \Cdot k_i=0$.
The linearised Riemann tensor of the on-shell particles then has the very simple form
\begin{align}
R_{i}^{\mu\nu\alpha\beta}={-}\left( \frac{\kappa}{2}\right) F_{i}^{\mu\nu} F_{i}^{\alpha\beta}
\ \ \text{with} \ \ F_{i}^{\mu\nu} = k_{i}^{\mu} \varepsilon_{k_i}^\nu - k_{i}^{\nu} \varepsilon_{k_i}^\mu \ ,\quad i=1,2\, , 
\end{align}
and, for the off-shell Riemann tensor, 
\begin{align}
  R_{3}^{\mu\nu\alpha\beta}= -\left( \frac{\kappa}{2}\right) \Big( 
  k_{3}^\mu k_{3}^\beta \, h_3^{\nu\alpha }+k_{3}^\nu
  k_{3}^\alpha
  \, h_3^{\mu \beta}-k_{3}^\nu k_{3}^\beta \, h_3^{\mu\alpha}-k_{3}^\mu k_{3}^\alpha \, h_3^{\nu\beta}
\Big)   \, ,
\end{align}
where $h_3^{\mu\nu}$ is an off-shell graviton with momentum $k_3$. The two vertices arising from an insertion of $I_1$ or $I_2$ can then be written in manifestly gauge invariant form~as%
\footnote{Our convention for symmetrisation is $A_{(\mu\nu)} = \frac{1}{2}(A_{\mu\nu}+A_{\nu\mu})$, and all momenta are assumed to be outgoing in this section.}
\begin{align}
\label{V1}
V_{I_1}^{\mu \nu}  (k_1,k_2) &= 
\raisebox{-.48\height}{
\begin{tikzpicture}[line width=1.5 pt,scale=.5]
	\draw[double, vector, line width = .75] (0:0) -- (0:2);
	\draw[double, vector, line width = .75] (0:0) -- (120:2);
	\draw[double, vector, line width = .75] (0:0) -- (240:2);
        \node at (60:1.2) {$I_1$};
        \node at (-90:2) {$1$};
        \node at (90:2) {$2$};
         \node at (0:2.6) {$\mu\nu$};
\end{tikzpicture}
}  = 
-i\, 3!\left( \frac{\kappa}{2}\right)^3\, 4 \,  \Tr (F_{1} \Cdot F_{2})  
(k_2 \Cdot F_{1})^{(\mu} (k_1 \Cdot F_{2})^{\nu )}
\ , 
\\ 
\label{V2}
V_{I_2}^{\mu \nu} (k_1,k_2) &= 
\raisebox{-.48\height}{
\begin{tikzpicture}[line width=1.5 pt,scale=.5]
	\draw[double, vector, line width = .75] (0:0) -- (0:2);
	\draw[double, vector, line width = .75] (0:0) -- (120:2);
	\draw[double, vector, line width = .75] (0:0) -- (240:2);
        \node at (60:1.2) {$I_2$};
        \node at (-90:2) {$1$};
        \node at (90:2) {$2$};
         \node at (0:2.6) {$\mu\nu$};
\end{tikzpicture}
} = 
i\, 3!\left( \frac{\kappa}{2}\right)^3\,  (k_3 \Cdot [F_{1}, F_{2}])^{\mu} (k_3 \Cdot [F_{1}, F_{2}])^{\nu}
\ . 
\end{align}
In the rest of the paper we will pick the gauge where $k_1\Cdot \varepsilon_{k_2} = k_2\Cdot \varepsilon_{k_1}= 0$. 
In this gauge, the two vertices are 
\begin{align}
V_{I_1}^{\mu \nu}  (k_1,k_2) &= 
\raisebox{-.48\height}{
\begin{tikzpicture}[line width=1.5 pt,scale=.5]
	\draw[double, vector, line width = .75] (0:0) -- (0:2);
	\draw[double, vector, line width = .75] (0:0) -- (120:2);
	\draw[double, vector, line width = .75] (0:0) -- (240:2);
        \node at (60:1.2) {$I_1$};
        \node at (-90:2) {$1$};
        \node at (90:2) {$2$};
         \node at (0:2.6) {$\mu\nu$};
\end{tikzpicture}
} \ = 
i\, 3!\left( \frac{\kappa}{2}\right)^3\, 8 (k_1\Cdot k_2)^3 \, ( \varepsilon_{k_1} \Cdot\varepsilon_{k_2}) 
 \varepsilon^{(\mu}_{k_1} \varepsilon_{k_2}^{\ \nu)}
\ , 
\\ 
V_{I_2}^{\mu \nu} (k_1,k_2) &= 
\raisebox{-.48\height}{
\begin{tikzpicture}[line width=1.5 pt,scale=.5]
	\draw[double, vector, line width = .75] (0:0) -- (0:2);
	\draw[double, vector, line width = .75] (0:0) -- (120:2);
	\draw[double, vector, line width = .75] (0:0) -- (240:2);
        \node at (60:1.2) {$I_2$};
        \node at (-90:2) {$1$};
        \node at (90:2) {$2$};
         \node at (0:2.6) {$\mu\nu$};
\end{tikzpicture}
}\ = 
i\, 3!\left( \frac{\kappa}{2}\right)^3\,  (\varepsilon_{k_1} \Cdot\varepsilon_{k_2})^2  (k_1\Cdot k_2)^2 (k_1 - k_2)^\mu (k_1 - k_2)^\nu  
\ \,.
\end{align}
In spinor-helicity variables the vertices become, for the 
$1^{++} 2^{++}$ helicity configuration,  
\begin{align}
V_{I_1}^{\mu \nu} (k_1^{++},k_2^{++}) &= -i\frac{3!}{4}\left( \frac{\kappa}{2}\right)^3\, [1\, 2]^4 \Big( \langle 2| \mu | 1] \langle 1 | \nu | 2]  + \mu \leftrightarrow \nu \Big) \, 
\ , 
\\ 
V_{I_2}^{\mu \nu} (k_1^{++},k_2^{++})&= i\frac{3!}{4}\left( \frac{\kappa}{2}\right)^3\,  [1\, 2]^4    (k_1 - k_2)^\mu (k_1 - k_2)^\nu  
\ , 
\end{align}
 while $V_{I_1} (k_1^{\pm\pm},k_2^{\mp\mp}) = V_{I_2} (k_1^{\pm\pm},k_2^{\mp\mp}) =0$.

In order to build the classical Compton amplitudes induced by the cubic interactions, we contract the three-point vertices above with%
\footnote{The factor of $-i$ is chosen in such a way that the spinless ($a^\mu=0$) limit of the corresponding amplitude matches the amplitude obtained from the three-point scalar-scalar-graviton vertex 
\begin{align}
\label{js}
V^{\mu \nu}_{\phi_m \phi_m h} (p_1, p_2) \ =
\raisebox{-.48\height}{
\begin{tikzpicture}[line width=1. pt, scale=.5]
	\draw[double, vector, line width = .75] (0:0) -- (0:2);
	\draw[thick,fermionnoarrow,  line width = .75] (0:0) -- (120:2);
	\draw[thick,fermionnoarrow,  line width = .75] (0:0) -- (240:2);
        \node at (-90:2) {$1^{\phi_m}$};
        \node at (90:2) {$2^{\phi_m}$};
         \node at (0:2.6) {$\mu\nu$};
\end{tikzpicture}
}
= \ i\left(\frac{\kappa}{2} \right)\Big[ -\eta^{\mu \nu} (p_1\Cdot p_2 + m^2) + p_1^\mu p_2^\nu + p_2^\nu p_1^\mu\Big]
\ , 
\end{align} 
where in the classical limit $p_1 \to p$ and $p_2 \to -p$.}
$-i T^{\mu \nu}$, where  $T^{\mu \nu}$ is the linearised energy-momentum tensor for a classical Kerr black hole of mass $m$, momentum $p^\mu$  and spin $S^\mu \coloneq m\,  a^
\mu$~\cite{Vines:2017hyw,Guevara:2018wpp,Chung:2019yfs,Bern:2020buy,Bautista:2021wfy},
\begin{align}
\label{Tmunu}
T^{\mu \nu} = \left( \frac{\kappa}{2} \right) 2 \left[ \cosh (a\Cdot q) p^\mu p^\nu  - i \frac{\sinh (a\Cdot q)}{a \Cdot q} p^{(\mu} \epsilon ^{\nu)}_{\ \alpha \beta \gamma}p^\alpha a^\beta q^\gamma\right]    \, , 
\end{align}
where
$q = k_1 + k_2$. We refer to  $a^\mu$ as the \textit{ring radius} vector~\cite{Cangemi:2022abk}. For a Kerr black hole in the rest frame and spinning around the $z$-axis we have $a^\mu = (0,0,0,a)$, where $a$ is the radius of the ring singularity; 
    the spin vector $S^\mu$
is related to the spin tensor~as
\begin{align}
\label{def:spinvector}
S^\mu = \frac{1}{2 m} \epsilon^{\mu\nu\alpha\beta} p_{\nu} S_{\alpha\beta} \ .
\end{align}
Note that  the ring radius satisfies 
\begin{align}
    p\Cdot a(p) =0\, . 
\end{align}  
    Using \eqref{Tmunu} and the standard de Donder propagator, 
    one obtains the following spinning Compton amplitudes: 
    \begin{align}
    \begin{split}
    \label{I1amplitude}
    M_{I_1}(p, k_1, k_2) &= i\,  \left( \frac{\kappa}{2}\right)^4   24\,(k_1 \Cdot k_2)^2 (\varepsilon_{k_1}\Cdot 
    \varepsilon_{k_2}) 
    \bigg\{ \cosh ( a\Cdot q ) \bigg[ 2 (p\Cdot \varepsilon_{k_1}) (p\Cdot \varepsilon_{k_2}) - m^2 (\varepsilon_{k_1} \Cdot\varepsilon_{k_2})  \Big]\\
 & -i \frac{\sinh (a\Cdot q)}{a \Cdot q}\Big[ 
(p\Cdot \varepsilon_{k_1} ) \epsilon( \varepsilon_{k_2} p a q) +(p\Cdot \varepsilon_{k_2} ) \epsilon( \varepsilon_{k_1} p a q) 
\Big] \bigg\} \, , 
  \end{split}
    \end{align}
    and
\begin{align}
    \begin{split}
    \label{I2amplitude}
    M_{I_2}(p, k_1, k_2) &=
    i  \,  \left( \frac{\kappa}{2}\right)^4 6 \, (k_1 \Cdot k_2) \,  (\varepsilon_{k_1}\Cdot 
    \varepsilon_{k_2})^2  
    \bigg\{\cosh ( a\Cdot q ) \Big[ \big( p\Cdot (k_1 - k_2)\big)^2 + m^2 (k_1 \Cdot k_2)   \Big]\\
 & -2i \, \frac{\sinh (a\Cdot q)}{a \Cdot q} p\Cdot (k_1 - k_2)\  \epsilon( k_1 k_2  p a)  
\Big] \bigg\} \, , 
  \end{split}
    \end{align}
    and we note that  
    \begin{align}
        M_{I_1}(p, k_1^{\pm\pm}, k_2^{\mp\mp})= M_{I_2}(p, k_1^{\pm\pm}, k_2^{\mp\mp})=0 
        \ . 
        \end{align}
The  second line of \eqref{I1amplitude} can be rewritten without Levi-Civita symbols using the relation
\begin{align}
   (p\Cdot \varepsilon_{k_1}^{\pm} ) \epsilon( \varepsilon_{k_2}^{\pm} p a q) +(p\Cdot \varepsilon_{k_2}^{\pm} ) \epsilon( \varepsilon_{k_1}^{\pm} p a q)  = \mp i p\Cdot (k_1 - k_2) \Big[ (p\Cdot \varepsilon_{k_1}^{\pm})(a\Cdot \varepsilon_{k_2}^{\pm})  -(p\Cdot \varepsilon_{k_2}^{\pm})(a\Cdot \varepsilon_{k_1}^{\pm}) \Big]\, , 
\end{align}
which can be proven via the identities
\begin{align}
\begin{split}
\label{LCidentities}
\epsilon( \varepsilon^{\pm}_{k_1} p a q) &= \mp i \Big[ a\Cdot(k_1 - k_2) (p\Cdot \varepsilon^{\pm}_{k_1}) - p\Cdot (k_1 - k_2) ( a\Cdot \varepsilon^{\pm}_{k_1})\Big]\ , \\
\epsilon( \varepsilon^{\pm}_{k_2} p a q) &= \pm i \Big[ a\Cdot(k_1 - k_2) (p\Cdot \varepsilon^{\pm}_{k_2}) - p\Cdot (k_1 - k_2) ( a\Cdot \varepsilon^{\pm}_{k_2})\Big]\ . 
\end{split}
    \end{align}
    In order to arrive at \eqref{LCidentities} we have used 
    \begin{align}
\varepsilon_{k_1}^{+}& = - \sqrt{2}\, \frac{\lambda_2\tilde{\lambda}_1}{\langle 1\, 2\rangle}\, , \qquad 
\varepsilon_{k_2}^{+} =  \sqrt{2}\, \frac{\lambda_1\tilde{\lambda}_2}{\langle 1\, 2\rangle}
\, , 
\\
\varepsilon_{k_1}^{-}& =  \sqrt{2}\, \frac{\lambda_1\tilde{\lambda}_2}{[1\, 2]}\, , \qquad \ \ \,
\varepsilon_{k_2}^{-} =  -\sqrt{2}\, \frac{\lambda_2\tilde{\lambda}_1}{[ 1\, 2]}\, , 
\end{align}
and the identity 
\begin{align}
        \label{LCidentity}
    \epsilon(abcd) = \frac{i}{4} \langle a|bcd - dcb |a]\, , 
\end{align}
valid if $a$ is a null vector.%
\footnote{We follow the   spinor conventions of  \cite{Brandhuber:2022qbk}.}
We can then further rewrite  \eqref{I1amplitude} and \eqref{I2amplitude}  in spinor-helicity form: 
\begin{align}
\begin{split}
\label{I1amplitude-bis}
M_{I_1}(p, k_1^{++}, k_2^{++}) &= 
i\left( \frac{\kappa}{2}\right)^4 3! \frac{[1\, 2]^4} {q^2}\bigg\{ - 4 \cosh(a\Cdot q)  \ (p\Cdot k_1)(p\Cdot k_2)\,    \\
 & +\frac{1}{2}p\Cdot(k_1-k_2) \frac{\sinh (a\Cdot q)}{a \Cdot q}\Big( [1|p|2\rangle [2|a|1\rangle - [2|p|1\rangle [1|a|2\rangle \Big) 
 \bigg\} \, , 
  \end{split}
    \end{align}
    and 
    \begin{align}
    \label{I2amplitude-bis}
        \begin{split}
            M_{I_2}(p, k_1^{++}, k_2^{++}) &= \frac{i}{2} \left( \frac{\kappa}{2}\right)^4 \, 3!\, \frac{[1\, 2]^4}{q^2}\bigg\{ \cosh (a\Cdot q) \Big[ \big( p\Cdot (k_1 - k_2)\big)^2 + m^2 (k_1 \Cdot k_2)   \Big]
            \\ 
            &+\frac{1}{2}p\Cdot(k_1-k_2) \frac{\sinh (a\Cdot q)}{a \Cdot q}\Big( [1|p|2\rangle [2|a|1\rangle - [2|p|1\rangle [1|a|2\rangle \Big)  \bigg\}\ , 
        \end{split}
    \end{align}
and similarly for the amplitudes with two negative-helicity  gravitons. 

Two comments are in order here. First,
we can now write the $G_3$ amplitude $M_{G_3} = M_{I_1} - 2 M_{I_2}$, getting
\begin{align}
\label{G3amplitude}
    M_{G_3}(p, k_1^{++}, k_2^{++}) = -3 \, i \, \left( \frac{\kappa}{2}\right)^4 \cosh (a\Cdot q) [1\, 2]^4 \left(m^2 + \frac{q^2}{2}\right)\, , 
\end{align}
and observe the remarkable feature that  it does not contain a term proportional to $\sinh (a\Cdot q) / a\Cdot q$. Also note that the $q^2/2$ term can be dropped in the classical limit. 
Second, in the spinless limit these expressions agree with those derived in \cite{Brandhuber:2019qpg}, up to a factor of $(-2/ \kappa^2)(\alpha^\prime/ 4)^2$ which was introduced  there to normalise  the   $I_1$ and $G_3$ couplings as they appear in the low-energy effective action of the bosonic string. In the spinless case, one has 
\begin{align}
   \begin{split}
    \left. M_{I_1}(p, k_1^{++}, k_2^{++})\right|_{a=0} &=
- 4i \left( \frac{\kappa}{2}\right)^4  3! \ \frac{[1\, 2]^4 }{q^2} (p\Cdot k_1)(p\Cdot k_2)\,   , \\ 
\left.M_{I_2}(p, k_1^{++}, k_2^{++})\right|_{a=0} &= \frac{i}{2}3! \left( \frac{\kappa}{2}\right)^4 \frac{[1\, 2]^4}{q^2}\Big[ \big( p\Cdot (k_1 - k_2)\big)^2 + m^2 (k_1 \Cdot k_2)   \Big]\ , 
\\
\left.M_{I_1}(p, k_1^{\pm\pm}, k_2^{\mp\mp})\right|_{a=0} &=\left.M_{I_2}(p, k_1^{\pm\pm}, k_2^{\mp\mp})\right|_{a=0}=0\ .
\end{split}
\end{align}

\section{One-loop spinning amplitudes from leading singularities}
\label{sec:1loopLS}

In this section we study the two-to-two scattering process of  Kerr black holes in the presence  of cubic deformations of gravity. There is no tree-level contribution to this process, hence the leading term requires a one-loop computation. 
Here we will perform this calculation using leading singularities, in two ways: first, using the   approach of \cite{Cachazo:2017jef,Guevara:2017csg}, slightly  adapted here to the spinning case;%
\footnote{See also \cite{Burger:2019wkq} for the case of two spinning/two non-spinning particles.}
we will then revisit the  computation using the  parameterisation of \cite{Bautista:2023szu}, which offers certain advantages compared to the previous one.

\subsection{Kinematics}
Our spinning particles have  masses $m_1$ and $m_2$  and spin vectors $a_1$ and $a_2$,   
\begin{equation}\label{eq: kinematics}
		\begin{tikzpicture}[scale=14,baseline={([yshift=-1mm]centro.base)}]
			\def\x{0}
			\def\y{0}

			\node at (0+\x,0+\y) (centro) {};
			\node at (-3pt+\x,-3pt+\y) (uno) {\footnotesize $p_1=\bar{p}_1 - \dfrac{q}{2}$};
			\node at (-3pt+\x,3pt+\y) (due) {\footnotesize $p_2=\bar{p}_2 + \dfrac{q}{2}$};
			\node at (3pt+\x,3pt+\y) (tre) {\footnotesize $p_2^\prime =  \bar{p}_2 - \dfrac{q}{2}$};
			\node at (3pt+\x,-3pt+\y) (quattro) {\footnotesize $\ \ p_1^\prime = \bar{p}_1 + \dfrac{q}{2}$};

			\draw [thick,double,orange] (uno) -- (centro);
			\draw [thick,double,amber] (due) -- (centro);
			\draw [thick,double,amber] (tre) -- (centro);
			\draw [thick,double,orange] (quattro) -- (centro);

			\draw [->] (-2.8pt+\x,-2pt+\y) -- (-1.8pt+\x,-1pt+\y);
			\draw [<-] (2.8pt+\x,-2pt+\y) -- (1.8pt+\x,-1pt+\y);
			\draw [<-] (-1.8pt+\x,1pt+\y) -- (-2.8pt+\x,2pt+\y);
			\draw [->] (1.8pt+\x,1pt+\y) -- (2.8pt+\x,2pt+\y);

			\shade [shading=radial] (centro) circle (1.5pt);
		\end{tikzpicture}
\end{equation}
with 
\begin{align}
 p_1^2 = (p_1^\prime)^2 = m_1^2\, , \qquad  p_2^2 = (p_2^\prime)^2 = m_2^2\, .
\end{align}
It is often convenient to introduce the  
barred variables \cite{Landshoff:1969yyn,Parra-Martinez:2020dzs}
\begin{align}
	\label{barredv}
	\begin{split}
		p_1 &= \bar{p}_1 - \frac{q}{2}\, , \qquad p_1^\prime = \bar{p}_1 + \frac{q}{2}\, , \\
		p_2 &= \bar{p}_2 +  \frac{q}{2}\, , \qquad p_2^\prime = \bar{p}_2 - \frac{q}{2}\, , 
	\end{split}
\end{align}
which satisfy  \begin{align}
	\label{eq: HEFTfame}
	\bar{p}_1\Cdot q =\bar{p}_2\Cdot q =0 \, ,  
\end{align}
and  $\bar{p}_i^2 = \bar{m}_i^2$, with 
\begin{equation}
\label{mbardef}
     \mb_i= \sqrt{m_i^2-\frac{q^2}{4}}=m_i -\frac{q^2}{4 m_i} + \cO(m_i^{-2})\, .
\end{equation}
We also introduce four-velocities $u_1$ and $u_2$ as
\begin{align}
    p_1 \coloneq m_1 u_1\, , \qquad  p_2 \coloneq m_2 u_2\, , 
\end{align}
with $u_1^2 = u_2^2 = 1$, as well as  the Lorentz factor
\begin{align}
\label{sigma}
\sigma \coloneqq  u_1 \Cdot u_2 \coloneq \frac{1}{\sqrt{1-v^2}} \ , 
\end{align}
where $v$ is the velocity of one particle in the rest frame of the other.
In the rest frame of particle $p_1$ the four-velocities have the form $u_1^\mu=(1,0,0,0)$ and $u_2^\mu=(\sigma,0,0,v \, \sigma)$.

\subsection{Introducing the triple cut}
\label{sect:triplecut}

\begin{figure}[H]
\centering
\begin{tikzpicture}[scale=1.5]
\def\size{0.3}
\begin{feynman}
    \vertex at (0,0) (top);
    \vertex at (-120:2) (L);
    \vertex at (-60:2) (R);
    \vertex at ($(L) + (180:1)$) [left](p4){\(p_1\)};
    \vertex at ($(R) + (0:1)$) [right](p3){\(p_1^\prime = p_1+q\)};
    \vertex at ($(top) + (180:1.5)$) [left](p1){\(p_2\)};
    \vertex at ($(top) + (0:1.5)$) [right](p2){\(p_2^\prime = p_2-q\)};
    \diagram*{
        (top) -- [momentum'={[arrow shorten=0.3]\(\ell_1\)}] (L);
        (top) -- [momentum={[arrow shorten=0.3]\(\ell_2\)}] (R);
        (L) -- [thick, momentum'={[arrow shorten=0.3]\(L=p_1+\ell_1\)}] (R);
        (p4) --[fermion, thick] (L);
        (R) --[fermion, thick] (p3);
        (p1) --[fermion,thick] (top);
        (top) --[fermion,thick] (p2);
    
    };
\end{feynman}

    \node (top) at (0,0) {};
    \node (L) at (-120:2) {};
    \node (R) at (-60:2) {};
    \draw[double, vector, line width = .75] (top) -- (L);
    \draw[double, vector, line width = .75] (top) -- (R);
    \path[fill=black!30] (top) circle [radius=\size, thick] node {\(M_{I}\)};
    \path[fill=black!30] (L) circle [radius=\size] node {\(M_3\)};
    \path[fill=black!30] (R) circle [radius=\size] node {\(M_3\)};
    \draw[thick,red, densely dashed] ($0.5*(L)+0.5*(R)+(0,0.2)$) -- ($0.5*(L)+0.5*(R)-(0,0.2)$);
    \draw[thick,red, densely dashed] ($0.5*(top)+0.5*(R)+(30:0.2)$) -- ($0.5*(top)+0.5*(R)+(-150:0.2)$);
    \draw[thick,red, densely dashed] ($0.5*(top)+0.5*(L)+(150:0.2)$) -- ($0.5*(top)+0.5*(L)+(-30:0.2)$);
\end{tikzpicture}
\caption{The leading singularity triangle diagram $\mathcal{M}^\bigtriangleup$, where  the bottom (top)  particles have mass $m_1$ ($m_2$). $M_I$ represents the Compton amplitude in the presence of a deformation $I$ with $I\in (I_1, I_2, G_3)$.}
\label{fig:LS}
\end{figure}
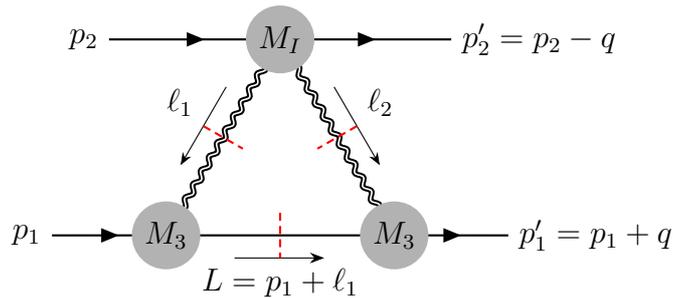

To begin, we note that the tree-level three-point amplitudes with two spinning particles and one graviton remains unchanged, hence the first $R^3$ corrections can only enter the four-point
spinning amplitude at one loop through the $R^3$ Compton amplitudes.
The relevant one-loop triple cut (leading singularity) is then  shown  in Figure~\ref{fig:LS}; there is  an additional diagram where the two types of particle are swapped. 
The top amplitude  is one of the  Compton amplitudes induced by the cubic interactions, as derived in Section~\ref{sec: ClassicalComptonAmplitudes}. 

The   massive-line cut in the figure    implies a sum over the polarisation states of the exchanged massive particle. This is usually realised by inserting a physical-state projector in between the two three-point amplitudes. For instance, consider the finite-spin three-point amplitudes given below, with two massive spin-$s$ states and one positive-helicity graviton, 
\begin{align}
\begin{split}
\label{3ptwithspin}
    &M_3 (-p_1, L, -\ell_1^{++}) = -i \kappa \frac{\braket{\bm{p_1 L}}^{2s}}{m_1^{2s}} (\ep_{-\ell_1}^{+}\Cdot p_1)^2\, , \qquad \\
    &M_3 (-L, p_1', -\ell_2^{++}) = -i \kappa \frac{\braket{\bm{L p_1'}}^{2s}}{m_1^{2s}} (\ep_{-\ell_2}^{+}\Cdot p_1')^2\, ,  
\end{split}
\end{align}
where the bold letters denote  massive spinors defined as 
\begin{align}
\label{defboldp}    
\ket{\bm{p}} \coloneq z_{a}^{(p)} \ket{p^a}\, , 
\end{align}
where $a = 1,2$ is the $SU(2)$ little-group index and $z_{a}^{(p)}$ is a polarisation variable. The massive factorisation cut in terms of these amplitudes is
\begin{equation}
\label{eq:QMmassivecut}
    - \kappa^2 \frac{(\braket{\bm{p_1} L_a} \braket{L^a \bm{p_1'}})^{2s}}{m_1^{4s}} (\ep_{-\ell_1}^{+}\cdot p_1)^2 (\ep_{-\ell_2}^{+}\cdot p_1')^2 =
    - \kappa^2 \frac{\braket{\bm{p_1} \bm{p_1'}}^{2s}}{m_1^{2s}} (\ep_{-\ell_1}^{+}\cdot p_1)^2 (\ep_{-\ell_2}^{+}\cdot p_1')^2,
\end{equation}
where $|L_a\rangle \langle L^a| = \epsilon_{a b} |L^b\rangle \langle L^a|$ is the state sum for the exchanged particle $L$, and $\epsilon_{a b}$ is the two-dimensional Levi-Civita tensor. Note that in arriving at the right-hand side of \eqref{eq:QMmassivecut} we have used $\braket{\bm{p_1} L_a} \braket{L^a \bm{p_1'}} = m_1 \braket{\bm{p_1} \bm{p_1'}}$. To compute the classical limit of \eqref{eq:QMmassivecut}, we use the methods outlined in \cite{Cangemi:2023ysz} to obtain
\begin{equation}
\label{eq:CLmassivecut}
    -\kappa^2 e^{\mp(\ell_1+\ell_2)\Cdot a_1} (\ep_{-\ell_1}^{\pm}\Cdot p_1)^2 (\ep_{-\ell_2}^{\pm}\Cdot p_1^\prime)^2 ,
\end{equation}
where $a_1$ is the spin vector of  particle 1. Note that we have included the case of two negative-helicity gravitons, obtained from \eqref{eq:QMmassivecut} by conjugation. Rewriting $e^{\mp(\ell_1+\ell_2)\Cdot a_1} = e^{\mp \ell_1 \Cdot a_1}e^{\mp \ell_2\Cdot a_1}$, we see that \eqref{eq:CLmassivecut} is just the product of two classical three-point amplitudes; hence, the state sum trivialises in the classical limit. Summarising, we can equivalently make use of the classical three-point amplitudes with one graviton of momentum $\ell$ and two massive particles with momenta $p$ and $-(p+\ell)$ and spin vector $a$, which are~\cite{Guevara:2018wpp}
\begin{align}
\label{3pc}
    M_3 (p, \ell^{++}) = -i \left( \frac{\kappa}{2}\right) e^{a\Cdot \ell} \left(
    \frac{\langle \xi | p | \ell]}{\langle \xi \ell\rangle }
    \right)^2\, , \qquad 
    M_3 (p, \ell^{--}) = -i \left( \frac{\kappa}{2}\right) e^{-a\Cdot \ell} \left(
    \frac{\langle \ell  | p | \xi ]}{[ \ell \xi] }
    \right)^2\, , \qquad 
\end{align}
where $\xi$ and $\tilde{\xi}$ are reference spinors. 

\subsection{Leading singularities with spinor parameterisation}
\label{1loopCG}

\subsubsection{Parameterisation of the momenta and integration measure}\label{sec: GuevaraParam}

In this approach,  one introduces  the following  parameterisation of the massive momenta $p_1$ and $p_1^\prime$ in Figure~\ref{fig:LS} \cite{Guevara:2017csg}, 
\begin{align}
     p_1 &=  \lambda \tilde{\eta} + \eta \tilde{\lambda}\, , \qquad p_1^\prime = \lambda \tilde{\lambda} + \beta \, \lambda \tilde{\eta} + \beta^{-1} \, \eta \tilde{\lambda}\ .
     \end{align}
     Then 
     \begin{align}
         p_1^2 = (p_1^\prime)^2 = \langle \lambda \eta\rangle [\tilde{\lambda} \tilde{\eta}] = m_1^2
     \, , 
     \end{align}
    which one can  solve by setting $\langle \lambda \eta\rangle  = [\tilde{\lambda} \tilde{\eta}]= m_1$.
The momentum transfer is 
\begin{align}
   q\coloneq   \ell_1+\ell_2 =  p_1^\prime - p_1 = \lambda \tilde{\lambda} + (\beta-1) \lambda \tilde{\eta} - \frac{\beta-1}{\beta}\eta \tilde{\lambda} \ , 
\end{align}
with  
\begin{align}
\label{qquadro}
    q^2  = - m_1^2 \frac{(\beta-1)^2}{\beta}\ ,  
\end{align}
hence the limit $q^2\to 0$ can be taken by sending $\beta \to 1$.
The loop momentum $L$ of Figure~\ref{fig:LS} can be parameterised as 
\begin{align}
    L = z \ell + \omega \lambda \tilde{\lambda}\, , 
\end{align}
where the null momentum $\ell$ can be expanded as 
\begin{align}
    \ell = A \, \lambda \tilde{\eta} + B \, \eta \tilde{\lambda} + A B \, \lambda \tilde{\lambda} + \eta\tilde{\eta}\ .  
\end{align}
After imposing the triple-cut conditions
\begin{align}
\label{triplecutcond}
    \ell_1^2 =\ell_2^2 = L^2-m_1^2 = 0\, ,
\end{align}
 there is only one integration variable left, which can be taken to be $y = - z / (\beta-1)^2$. 
The cut loop momenta $\ell_1$ and $\ell_2$ can then be  solved as~\cite{Guevara:2017csg} 
\begin{align}
\begin{split}
    \ell_1 =& \frac{1}{\beta+1}\Big[ \frac{1 + \beta y}{y }\lambda - (\beta^2-1) \eta\Big]\times \frac{1}{\beta+1}\Big[(1 + \beta y ) \tilde{\lambda} + y (\beta^2 -1) \tilde{\eta}\Big]\, , 
    \\ 
     \ell_2=& \frac{i}{\beta+1}\Big[ \frac{1 -y}{y }\lambda + \frac{\beta^2-1}{\beta} \eta\Big]\times \frac{i}{\beta+1}\Big[(1 - \beta^2 y ) \tilde{\lambda} - y \beta (\beta^2 -1) \tilde{\eta}\Big]\, .
\end{split} 
\end{align}
The one-loop integration measure,
\begin{align}
    \frac{d^4 L }{(2\pi)^4}\frac{1}{\ell_1^2}\frac{1}{\ell_2^2}
    \frac{1}{L^2- m_1^2}\ ,
\end{align}
then becomes, after integrating out $\omega$, $A$ and $B$,  
\begin{align}
 d\mathfrak{m}_{\rm eff}= \frac{1}{32\pi m_1^2}\frac{\beta}{\beta^2-1}\ \frac{dy}{y}\ . 
\end{align}
We have   normalised this effective measure  in such a way that by integrating it and picking the residue at $y=\infty$ one gets the expected result for the classical part of a  scalar triangle, that is%
\footnote{Here we have exploited  the nice fact that the classical part of the triangle integral is simply $-1/4$ times the result of its leading singularity \cite{Cachazo:2017jef,Guevara:2018wpp}.}
\begin{align}
 \ointctrclockwise_{y= \infty}\!d\mathfrak{m}_{\rm eff} =   2 \pi i \,  {\rm Res} \left[ \frac{1}{32\pi m_1^2}\frac{\beta}{\beta^2-1}\ \frac{1}{y} \right]_{y=\infty}  = - \frac{i}{32\,  m_1  \sqrt{ -q^2} } + \cO( 1)\ , 
\end{align}
after using \eqref{qquadro}. 

\subsubsection{Taking the  classical limit}

In the absence of spin, the classical limit can simply be taken by expanding around $q^2=0$, or equivalently $\beta=1$ (this is known as the holomorphic classical limit, or HCL \cite{Guevara:2017csg}). In the presence of spin, 
the correct classical limit can be obtained by recalling that $q$ and $a$ scale with $\hbar$ as 
\begin{align}
\label{hbarscaling}
    (q, a)\to (\hbar q, \hbar^{-1} a)\ , 
    \end{align}
and then expanding in $\hbar$.
The result of the leading singularity calculation is conveniently expressed in terms of the 
following variables: 
\begin{align}\label{eq: GuevaraVariables}
\begin{split}
    u &\coloneqq 
    \langle \eta | p_2 | \lambda] \, , \qquad 
v \coloneqq  \langle \lambda | p_2 | \eta]  
\, ,  \\
 w &\coloneqq \langle \eta | a| \lambda]\, ,  \qquad \ 
z \coloneqq \langle \lambda | a| \eta]\, , 
\end{split}
\end{align}
where
\begin{align} \label{eq: uvExpressions}
    u+v  = 2 p_1\Cdot p_2  \coloneq  2 m_1 m_2 \, \sigma
    \, , \qquad
    u \, v = m_1^2 m_2^2 + \langle \eta|p_1 |\eta] \langle \lambda|p_1 |\lambda]\, ,  
\end{align}
where
\begin{align}
    \langle \lambda|p_2|\lambda] = 2 (q\Cdot p_2) + (\beta-1) \Big( \frac{u}{\beta}-v\Big)\, , \qquad  
    \langle\lambda|a|\lambda] = 2 (q\Cdot a) + (\beta-1) \Big( \frac{w}{\beta}-z\Big)\, ,  
\end{align}
and $\sigma$ was introduced in \eqref{sigma}. 
We can then implement the rescaling \eqref{hbarscaling} on these variables~as
\begin{align}\label{eq: hbarScalingExtended}
    \begin{split}
(\lambda, \tilde{\lambda}) &\to \hbar^{\frac{1}{2}}(\lambda, \tilde{\lambda})\, , \quad (\eta, \tilde{\eta}) \to \hbar^{-\frac{1}{2}}(\eta, \tilde{\eta})\, , \quad 
p\Cdot q\to \hbar^2 p\Cdot q\, , 
\\
 (u, v) &\to (u, v)\, , \qquad (w, z) \to \hbar^{-1} (w, z) \, , \quad 
\beta-1 \to
   \hbar (\beta-1)  \, .  
\end{split}
\end{align}
An important comment is in order here. Simply  counting  powers of $\beta{-}1$ would discard terms such as $(\beta{-}1)\langle \lambda | p_2 | \lambda ] \langle \eta|a|\eta]$ compared to  terms such as  $u\langle \lambda|a|\lambda]$, though they scale identically in $\hbar$. Expanding in $\hbar$ (and not  in $\beta{-}1$) we can then  capture fully the classical contribution. Related issues of the $\beta \to 1$ prescription in the presence of spin are discussed in \cite{Bautista:2023szu}.

\subsubsection{Leading singularities for \texorpdfstring{$I_1$}{I1}}

There are two triple cuts to consider, where the helicities of the internal gravitons are $(++, ++)$ or $(--, --)$. There is no $(++, --)$ internal helicity assignment as the corresponding Compton amplitude vanishes. 
The leading singularities for these two internal helicity assignments are given by  
\begin{align}
\begin{split}
\label{twotriplecuts}
     \cM_{I_1}^{\bigtriangleup, (I)} &= 
     \ointctrclockwise_{y= \infty}\!d\mathfrak{m}_{\rm eff}\ M_{I_1}(p_2, \ell_1^{++}, \ell_2^{++}) M_3 (p_1^\prime , -\ell_2^{--}) M_3 (-p_1, -\ell_1^{--})\, , 
    \\
     \cM_{I_1}^{\bigtriangleup, (II)} &=
     \ointctrclockwise_{y= \infty}\!d\mathfrak{m}_{\rm eff}\ M_{I_1}(p_2, \ell_1^{--}, \ell_2^{--}) M_3 (p_1^\prime , -\ell_2^{++}) M_3 (-p_1, -\ell_1^{++})\, , 
     \end{split}
\end{align}
which requires the cubic gravity Compton amplitude given in \eqref{I1amplitude-bis}, its parity conjugate expression, and  the three-point amplitudes  in \eqref{3pc}. 
Note that the exponential factors contained in these three-point amplitudes combine to 
$e^{\pm a_1\Cdot (\ell_1 + \ell_2) }= e^{\pm a_1\Cdot q}$ in the first/second triple cut in \eqref{twotriplecuts}. 

The triple cuts in \eqref{twotriplecuts}
will now be   evaluated on the cut using the parameterisation described in Section~\ref{sec: GuevaraParam}.
Any instance of the internal graviton momenta $\ell_1$ and $\ell_2$ can be written in terms of the $\lambda$ and $\eta$ spinors, meaning everything can be expressed using the spinor traces \eqref{eq: GuevaraVariables}. For instance,
\begin{equation}
    \langle \ell_1 \ell_2\rangle = i\frac{\beta-1}{\beta}\frac{m_1}{y}\, , \qquad [\ell_1 \ell_2] = -i(\beta-1)y\, m_1\, .
\end{equation}
This highlights the fact that angle and square brackets are not simply related by complex conjugation, since we are working with complex momenta.

It turns out that  the integrands (and hence leading singularities) for the two internal helicity assignments are identical, hence the two exponential factors combine as 
$e^{a_1\Cdot q}+e^{-a_1\Cdot q} =  2\cosh{(a_1\Cdot q)}$. 
We then  take the residue at $y=\infty$, and take the leading-in-$\hbar$ contribution using  the scalings \eqref{eq: hbarScalingExtended}. The first contribution is at $\cO(\hbar^3)$, for which we get\footnote{We used the SpinorHelicity4D package \cite{AccettulliHuber:2023ldr} for some spinor manipulations.}
\begin{align}
    \cM_{I_1}^\bigtriangleup = \cosh (a_1 \Cdot q)\Big( \cM_{I_1}^\bigtriangleup |_{\cosh{}} + 
    \cM_{I_1}^\bigtriangleup |_{\sinh{}}\Big)\, , 
\end{align}
with 
\begin{align}
    \cM_{I_1}^\bigtriangleup |_{\cosh{}} ={}& i\left(\frac{\kappa}{2}\right)^6\frac{3}{64}m_1^4 (\beta-1)^3\left[ (u+v)^2-4m_1^2m_2^2\right]\cosh{(a_2\Cdot q)}\, , \label{eq: GuevaraI1CoshTerm}\\
    \cM_{I_1}^\bigtriangleup |_{\sinh{}} ={}& - i\left(\frac{\kappa}{2}\right)^6\frac{3}{128}m_1^4 (\beta-1)^3 \frac{\sinh{(a_2\Cdot q)}}{a_2\Cdot q}(u+v) \Big[(u-v)\langle \lambda|a_2|\lambda] \notag\\  +& (z-w)\langle \lambda|p_2|\lambda]
+2(\beta-1)\Big(\langle\eta|p_2|\eta]\langle\lambda|a_2|\lambda]-\langle\eta|a_2|\eta]\langle\lambda|p_2|\lambda]\Big)\Big]\, . \label{eq: GuevaraI1SinhTerm}
\end{align}
\black
Notice that here it is apparent that the classical $\hbar$ expansion does not match the $(\beta-1)$ expansion when spin is included, as the $\sinh$ term contains  both a term proportional to  $(\beta-1)^3$ and $(\beta-1)^4$, which however are of the same order in $\hbar$. 

The $\sinh$ term can be simplified using the identity
\begin{align}\label{eq: EpsilonExpSH} 
   - 4i\epsilon(p_1 p_2 a_2 q) ={}&  (u-v)\langle\lambda|a_2|\lambda] + (z-w)\langle\lambda|p_2|\lambda] \notag \\
    & + 2 (\beta-1)\Big(\langle\eta|p_2|\eta]\langle\lambda|a_2|\lambda]-\langle\eta|a_2|\eta]\langle\lambda|p_2|\lambda]\Big) + \cO(\hbar)\, , 
    \black
\end{align}
which can be derived using \eqref{LCidentity}. 
Using then  \eqref{eq: uvExpressions} to rewrite the remaining spinor traces  in terms of familiar kinematic variables, we can recast the $\cM_{I_1}^\bigtriangleup$ leading singularity in the compact form 
\begin{equation}
\label{I1amp}
    \cM_{I_1}^\bigtriangleup = i \left(\frac{\kappa}{2}\right)^6\frac{3}{16}  m_1^3 m_2^2
    \, |q|^{3}
    \cosh(a_1\Cdot q)\left[\left(\sigma^2-1\right) \cosh(a_2\Cdot q) -  i \sigma   \frac{\sinh(a_2\Cdot q)}{a_2\Cdot q}\epsilon (u_1 u_2 a_2 q)\right]\, ,
\end{equation}
where we have defined 
 \begin{align}
 \label{qdef}
     |q| \coloneq 
 \sqrt{-q^2}\, , 
 \end{align}
 with  $q^2{<}0$.
Finally, to obtain the complete one-loop result we need to add a term $ \cM_{I_1}^\bigtriangledown$ obtained from 
$ \cM_{I_1}^\bigtriangleup$ by performing the the simultaneous replacements%
\footnote{One should  also swap  $u_1\leftrightarrow u_2$  
and $q\leftrightarrow -q$ but these two changes compensate each other.}
$m_1\leftrightarrow m_2$ and  $a_1\leftrightarrow a_2$, 
\begin{align}
\label{swapped}
   \cM_{I_1}^\bigtriangledown \coloneq \left.\cM_{I_1}^\bigtriangleup \right|_{m_1\leftrightarrow m_2, a_1\leftrightarrow a_2}\, . 
\end{align}
Doing so we arrive at the final answer 
\begin{align}
\label{completeamp}
 \cM_{I_1} = \cM_{I_1}^\bigtriangleup  +\cM_{I_1}^\bigtriangledown\, , 
\end{align}
for the one-loop classical amplitude. This quantity will be the main ingredient to compute the momentum and spin kicks in Section~\ref{sec:Obs-from-KMOC}.

\subsubsection{Leading singularities for \texorpdfstring{$I_2$}{I2} and \texorpdfstring{$G_3$}{G3}}
The calculation for the $I_2$ and $G_3$ leading singularities is very similar to the case of $I_1$, with the relevant cubic gravity Compton amplitudes for the $I_2$ and $G_3$ cases  now being given in \eqref{I2amplitude-bis} and \eqref{G3amplitude}, respectively. Proceeding as before, we arrive at the following results: 
\begin{align}
\label{I2amp}
    \cM_{I_2}^\bigtriangleup &= i \left(\frac{\kappa}{2}\right)^6\frac{3}{32}  m_1^3 m_2^2
    \, |q|^{3}
    \cosh(a_1\Cdot q)\left[\sigma^2 \, \cosh(a_2\Cdot q) - i \sigma  \frac{\sinh(a_2\Cdot q)}{a_2\Cdot q}\epsilon (u_1 u_2 a_2 q)\right]\, , 
\\
\label{G3amp}
    \cM_{G_3}^\bigtriangleup &=- i \left(\frac{\kappa}{2}\right)^6\frac{3}{16}  m_1^3 m_2^2
    \, |q|^{3}
    \cosh(a_1\Cdot q)\, \cosh(a_2\Cdot q) \, . 
\end{align}
Note the interesting feature of the classical one-loop amplitude for the $G_3$ case, which is free of a $\sinh$ term (as the Compton amplitude \eqref{G3amplitude}). Furthermore, we can obtain the $G_3$ amplitude from the $I_1$ amplitude by (formally) setting ${\sigma}\rightarrow 0$.

Finally, as in the $I_1$ case,  to obtain the complete one-loop result we need to add terms $\cM_{I_2}^\bigtriangledown$, $\cM_{G_3}^\bigtriangledown$ obtained by   simultaneously replacing 
$m_1\leftrightarrow m_2$, $a_1\leftrightarrow a_2$ in $\cM_{I_2}^\bigtriangleup$, $\cM_{G_3}^\bigtriangleup$, so that the complete classical one-loop amplitudes are
\begin{align}
\label{completeamp2}
    \cM_{I_2} = \cM_{I_2}^\bigtriangleup  +\cM_{I_2}^\bigtriangledown\, ,
\qquad 
\cM_{G_3} = \cM_{G_3}^\bigtriangleup  +\cM_{G_3}^\bigtriangledown\, .
\end{align}
The results for $I_1$ in \eqref{I1amp} and $I_2$ in \eqref{I2amp} in the spinless limit agree with  \cite{Brandhuber:2019qpg,Emond:2019crr}. In addition, in the limit where $a_2\rightarrow0$ and $m_1\gg m_2$ 
(i.e.~in the probe limit) we match the results of \cite{Burger:2019wkq}.

\subsection{Leading singularities without spinor variables}

In this section we revisit the leading singularity computation of Section~\ref{1loopCG} using an efficient, alternative parameterisation of \cite{Bautista:2023szu} that avoids the use of auxiliary spinor variables to solve the triple-cut on-shell conditions.
We  begin by  introducing  a convenient linear combination $\tilde{p}_2$ of $p_1$ and $p_2$, 
\begin{align}
\begin{split}
\tilde{p}_2^\mu & = p_2^\mu - \frac{m_2}{m_1} \sigma\,  p_1^\mu = m_2 (u_2^\mu - \sigma\,  u_1^\mu) =
m_2 \tilde{u}_2^\mu \ , \\
    \end{split}
\end{align} 
 which  is orthogonal to $p_1$, i.e.~$\tilde{p}_2 \Cdot p_1=0$.
It is convenient to parameterise the internal graviton momentum  $\ell_1$ as
\begin{align}
    \ell_1^\mu = \alpha |q| \tilde{u}_2^\mu + \beta |q| u_1^\mu + \gamma q^\mu + i \delta E^\mu \ ,
\end{align}
where $E^\mu = \epsilon^{\mu \alpha \beta \gamma} \tilde{u}_{2\alpha} u_{1\beta} q_\gamma$, with $|q| $ being defined in \eqref{qdef}. In the first step we perform three residue integrals corresponding to the triple cut loci conditions \eqref{triplecutcond}, which leads to the following
one-parameter family of solutions:
\begin{align}\label{bestpara}
\begin{split}
\ell_1^\mu & =  \frac{q^\mu}{2}  + \frac{\sqrt{z^2-1}}{2 v \sigma} |q| \, \tilde{u}_2^\mu + \frac{i z}{2 v \sigma} E^\mu \ , \\
\ell_2^\mu & =  \frac{q^\mu}{2}  - \frac{\sqrt{z^2-1}}{2 v \sigma} |q| \, \tilde{u}_2^\mu - \frac{i z}{2 v \sigma} E^\mu \ ,
\end{split}
\end{align}
where the parameter $z$ is related to the parameter $y$ in Section~\ref{1loopCG} via $z=\frac{1+y^2}{2 y}$. The remaining integration measure is proportional to $\frac{1}{\sqrt{z^2-1}}$ where the precise normalisation comes from Jacobian factors arising at the various steps of the computation. For the evaluation of the contour integrals it is useful to quote here its residue in the parameterisation given above:
\begin{align}
{\underset{z=\infty}{\mathrm{Res}}\,} \frac{1}{\sqrt{z^2-1}} = -1\, .
\end{align}
The rest of the computation of the leading singularities
is very similar to that of \eqref{twotriplecuts} in the previous section. In the present case the effective measure is 
\begin{align}
\label{dmeff2}
    d\mathfrak{m}_{\mathrm{eff}} = \frac{1}{64 \pi m_1 |q|} \frac{dz}{\sqrt{z^2-1}}\, ,
\end{align}
where the contour has to be taken around $z=+\infty$
and, as before, the normalisation is chosen to give directly the 
classical contribution from the triangle integral. The two triple cuts are then
\begin{align}
\begin{split}
\label{twotriplecuts-bis}
     \cM_{I}^{\bigtriangleup, (I)} &=  
     \ointctrclockwise_{z= +\infty} 
     \!d\mathfrak{m}_{\mathrm{eff}} \ M_{I}(p_2, \ell_1^{++}, \ell_2^{++}) M_3 (p_1^\prime , -\ell_2^{--}) M_3 (-p_1, -\ell_1^{--})  \, , 
    \\
     \cM_{I}^{\bigtriangleup, (II)} & =  
     \ointctrclockwise_{z= +\infty} 
     \!d\mathfrak{m}_{\mathrm{eff}} \ M_{I}(p_2, \ell_1^{--}, \ell_2^{--}) M_3 (p_1^\prime , -\ell_2^{++}) M_3 (-p_1, -\ell_1^{++})  \, ,
     \end{split}
\end{align}
with $I{\in} (I_1, I_2, G_3)$,  and   $d\mathfrak{m}_{\mathrm{eff}}$  given in \eqref{dmeff2}. Also note that, as in \eqref{twotriplecuts}, the first/second triple cut contains a factor of 
$e^{\pm a_1\Cdot q}$ from the two three-point amplitudes.

We will not present full details of the evaluation of these residues, but make a number of comments. First, the first term, $q^\mu/2$, in \eqref{bestpara} never contributes as it either contracts to zero or gives terms like $p_{1,2} \Cdot q \propto q^2$ which can be dropped. Similarly, terms involving $E^\mu$ contract to zero or give rise to terms with vanishing residue. So effectively we can set
\begin{align}
    \ell_{1,2} \rightarrow \pm \frac{\sqrt{z^2-1}}{2 v \sigma} |q| \, \tilde{u}_2^\mu \ ,
\end{align}
and it is easy to see that only terms quadratic in $\ell_1$ can contribute.
The corresponding residue is
\begin{align}
{\underset{z=\infty}{\mathrm{Res}}\,} \sqrt{z^2-1} = +\frac{1}{2} \ .
\end{align}
Plugging the tree amplitudes from previous sections into the residue integral and performing the state sum we arrive at the following results for the classical amplitudes
(setting the couplings of $I_1$ and $G_3=I_1 -2 I_2$ to unity), 
\begin{align}
\label{triple-cut-result}
\begin{split}
    \cM_{I_1}^\bigtriangleup  & = i \left(\frac{\kappa}{2}\right)^6 \frac{3}{16}  m_1^3 m_2^2 |q|^3 
    \cosh (a_1 \Cdot q)
    \left[ (\sigma^2-1) \cosh (a_2 \Cdot q) - i \sigma \frac{\sinh (a_2 \Cdot q)}{a_2 \Cdot q}
    \epsilon(u_1 u_2 a_2 q)\right]  , \\
    \cM_{G_3}^\bigtriangleup  & = -i \left(\frac{\kappa}{2}\right)^6   \frac{3}{16}   m_1^3 m_2^2 |q|^3  \cosh (a_1 \Cdot q)
     \cosh (a_2 \Cdot q)  , 
    \end{split}
\end{align}
in agreement with \eqref{I1amp} and   \eqref{G3amp}.
For the bosonic string one obtains%
\footnote{Recall that the particular values of $\beta_1$ and $\beta_2$ in bosonic string theory are given in \eqref{beta1beta2}.}
\begin{align}
\begin{split}
    \cM_{I_1, {\rm BS}}^\bigtriangleup  & = -i \frac{(\alpha^\prime)^2 \kappa^4 m_1^3 m_2^2}{2^{13}} |q|^3 \cosh (a_1 \Cdot q)
    \left[ (\sigma^2-1) \cosh (a_2 \Cdot q) - i \sigma \frac{\sinh (a_2 \Cdot q)}{a_2 \Cdot q}
    \epsilon(u_1 u_2 a_2 q)\right] , \\
    \cM_{G_3, {\rm BS}}^\bigtriangleup & = i \frac{(\alpha^\prime)^2 \kappa^4 m_1^3 m_2^2}{2^{12}} |q|^3 \cosh (a_1 \Cdot q)
     \cosh (a_2 \Cdot q) \ .
    \end{split}
\end{align}


\section{One-loop  spinning amplitudes from unitarity and the heavy-mass limit}
\label{sec:HEFT}

\subsection{Review of the heavy-mass limit for Einstein-Hilbert}

In this section we will cover an alternative method to construct the classical amplitudes above based on unitarity cuts and the heavy-mass limit. The discussion here follows the approach of Heavy-mass Effective Field Theory (HEFT)  \cite{Georgi:1990um,Luke:1992cs,Neubert:1993mb,Manohar:2000dt,Damgaard:2019lfh,Aoude:2020onz,Haddad:2020tvs,Brandhuber:2021kpo,Brandhuber:2021eyq,Brandhuber:2021bsf}, where the classical limit is taken by expanding in the $m{\rightarrow} \infty$ limit. This is equivalent to the usual $\hbar{\rightarrow} 0$  expansion for spinless particles since the typical graviton momentum $q$ is much less than the mass $m$ of the particle. 

To impose this hierarchy of scales one can either work with classical wavenumbers $\hat{q}=q/\hbar$ and then take $\hbar$ small with fixed $\hat{q}$, or simply take $m$ large compared to the exchanged graviton momentum. The same logic extends to the case of spinning particles, except that one has to simultaneously expand in the limit of $m{\rightarrow}\infty$ and $s{\rightarrow} \infty$ (where $s$ is the spin quantum number of the massive particle) while keeping the ratio $s/m=|a|$ finite. In practice, it is more convenient to work with the barred variables of \eqref{barredv} (since they satisfy $\pb_i\Cdot q=0$ exactly) and perform the expansion as $\mb{\rightarrow}\infty$. 
Once we have obtained the one-loop four-point classical amplitudes, we can safely remove the bars from our expressions as discussed later in this section. 

We begin with a review of the heavy-mass expansion in the case of pure general relativity coupled to scalars, before including $R^3$ corrections and spin. 
The one-loop four-point amplitude can be constructed by performing a unitarity cut in the $q^2$-channel and gluing together two Compton amplitudes. This extracts the piece of the amplitude containing discontinuities in $q^2$, which contribute to long-range effects after performing a Fourier transform in $q$. The analytic terms in $q$ would only give rise to local contributions in impact parameter space. To obtain the classical piece of this cut we must expand in the classical limit (or equivalently the heavy-mass limit) and extract the piece with the right scaling. Since we are already working at the level of the $q^2$ cut we can perform the classical expansion on each Compton amplitude separately. 

To warm up, we will first consider the case of the usual scalar Compton amplitude without $R^3$ corrections
\begin{align}
	\label{eq: comptonExpansion}
	M_{4}  = M_{4,\bar{m}^3} + M_{4,\bar{m}^2} + \cO(\mb^1),
\end{align}
where $M_{4,\bar{m}^3}$ and $M_{4,\bar{m}^2}$ are the hyperclassical and classical pieces of the Compton amplitude respectively. These are given explicitly by 
\begin{align}
	\label{eq: comptonheftm3}
	M_{4,\bar{m}^3} &\coloneqq  \cA_3(\pb,k_1)\,\pi\delta(\pb\Cdot k_1)\,\cA_3(\pb,k_2)\,,\quad \cA_{3}(\pb,k_i)\coloneqq -i\kappa (\pb\Cdot \varepsilon_{k_i})^2
	\\ \cr
	\label{eq: comptonheftm2}
	M_{4,\bar{m}^2}     & \coloneqq\cA_{4}(\pb,k_1,k_2)=    - i  \frac{
\kappa^2}{q^2}
	\left(\frac{\bar{p} \Cdot F_1 \Cdot F_2 \Cdot \bar{p}}{\bar{p}\Cdot k_1}\right)^2
	\, .
\end{align}
Note that the ``hyperclassical'' piece of the Compton amplitude is given by an iteration of three-point amplitudes. To compute the classical piece of the one-loop amplitude we will only need the first two terms in the heavy-mass expansion of the Compton amplitude. However, other classical observables may require further subleading terms, see for example \cite{Brunello:2024ibk}. 

We will write the terms in the expansion above diagrammatically as follows
\begin{align}
\label{eq: comptonExpansionDiagram}
	\begin{tikzpicture}[baseline={([yshift=-0.8ex]current bounding box.center)}]\tikzstyle{every node}=[font=\small]
		\begin{feynman}
			\vertex (a) {\(p\)};
			\vertex [right=1.8cm of a] (f2)[HV]{$M_4$};
			\vertex [right=1.8cm of f2] (c){$p'$};
			\vertex [above=1.8cm of f2] (gm){};
			\vertex [left=0.8cm of gm] (g2){$k_{1}$};
			\vertex [right=0.8cm of gm] (g20){$k_{2}$};
			\diagram* {
            (g2) -- [rmomentum'={}](f2), (g20) -- [rmomentum={}](f2),
			(a) -- [fermion,thick] (f2) --  [fermion,thick] (c),
			(g2)--[double, vector, line width = .75](f2),(g20)--[double, vector, line width = .75](f2)
			};
		\end{feynman}
	\end{tikzpicture}
	{=}
	\begin{tikzpicture}[baseline={([yshift=-0.8ex]current bounding box.center)}]\tikzstyle{every node}=[font=\small]
		\begin{feynman}
			\vertex (a) {\(p\)};
			\vertex [right=1.8cm of a] (f2)[HV]{H};
			\vertex [right=1.8cm of f2] (c)[HV]{H};
			\vertex [right=1.8cm of c] (d){\(p'\)};
			\vertex [above=1.8cm of f2] (g1){$k_{1}$};
			\vertex [above=1.8cm of c] (g2){$k_{2}$};
			\vertex [right=0.9 cm of f2] (cut);
			\vertex [above=0.3cm of cut] (cutu);
			\vertex [below=0.3cm of cut] (cutb);
			\diagram* {
            (g1) -- [rmomentum'={}] (f2), (g2) -- [rmomentum={}] (c),
			(a) -- [fermion,thick] (f2) --  [thick] (c)--  [fermion,thick] (d),
			(g1)--[double, vector, line width = .75](f2),
			(g2)--[double, vector, line width = .75](c),
			(cutu)--[red,thick] (cutb)
			};
		\end{feynman}
	\end{tikzpicture}
	{+}
	\begin{tikzpicture}[baseline={([yshift=-0.8ex]current bounding box.center)}]\tikzstyle{every node}=[font=\small]
		\begin{feynman}
			\vertex (a) {\(p\)};
			\vertex [right=1.8cm of a] (f2)[HV]{H};
			\vertex [right=1.8cm of f2] (c){$p'$};
			\vertex [above=1.8cm of f2] (gm){};
			\vertex [left=0.8cm of gm] (g2){$k_{1}$};
			\vertex [right=0.8cm of gm] (g20){$k_{2}$};
			\diagram* { (g2) -- [rmomentum'={}](f2), (g20) -- [rmomentum={}](f2),
			(a) -- [fermion,thick] (f2) --  [fermion,thick] (c),
			(g2)--[double, vector, line width = .75](f2),(g20)--[double, vector, line width = .75](f2)
			};
		\end{feynman}
	\end{tikzpicture}
 {+\cdots}
\end{align}
where the red cut propagator denotes the delta function $\pi \delta(\pb\Cdot k_1)$. 

To compute the classical part of the scalar four-point amplitude we need the terms in the heavy-mass expansion which are $\cO(\mb_1^3\mb_2^2)$ and $\cO(\mb_1^2\mb_2^3)$. To obtain these we simply glue one copy of \eqref{eq: comptonheftm3} and \eqref{eq: comptonheftm2} together in a unitary cut 
\begin{align}\label{eq: scalarEHFourPoint}
    &\begin{tikzpicture}[baseline={([yshift=-0.4ex]current bounding box.center)}]\tikzstyle{every node}=[font=\small]
            \node (b1) at (1,0) {};
            \node (b3) at (3,0) {};
            \node (u1) at (2,2.5) {};
            \draw[double, vector, line width = .75] (b1) -- (u1);
            \draw[double, vector, line width = .75] (b3) -- (u1);
		      \begin{feynman}
			\vertex (p1) {\(p_1\)};
			\vertex [above=2.5cm of p1](p2){$p_2$};
			\vertex [right=2cm of p2] (u1) [HV]{H};
			\vertex [right=2cm of u1] (p3){$p_{2'}$};
			\vertex [right=1.0cm of p1] (b1) [HV]{H};
			\vertex [right=1.0cm of b1] (b2) []{};
			\vertex [right=1.0cm of b2] (b3) [HV]{H};
			\vertex [right=1.0cm of b3](p4){$p_{1'}$};
			\vertex [above=1.25cm of p1] (cutL);
			\vertex [right=4.0cm of cutL] (cutR){};
			\vertex [right=1.0cm of b1] (cut1);
			\vertex [above=0.3cm of cut1] (cut1u);
			\vertex [below=0.3cm of cut1] (cut1b);
			\vertex [right=0.5cm of b2] (cut2);
			\vertex [above=0.3cm of cut2] (cut2u);
			\vertex [below=0.3cm of cut2] (cut2b);
			\diagram* {
			(p2) -- [thick] (u1) -- [thick] (p3),
			(b1)--[opacity=0,rmomentum=\(\ell_{1}\)](u1),
            (b3)-- [opacity=0,rmomentum'=\(\ell_{2}\)] (u1),
            (p1) -- [thick] (b1)-- [thick] (b3)-- [thick] (p4), (cutL)--[dashed, red,thick] (cutR), (cut1u)--[ red,thick] (cut1b),
			};
		\end{feynman}
	      \end{tikzpicture}\nn\\
       &=\frac{1}{2}\sum_{h_1,h_2}\int\frac{d^D\ell_1}{(2\pi)^D}\delta(\ell_1^2)\delta(\ell_2^2)\cA_3^{h_1}(\pb_1,\ell_1)\pi\delta(\pb_1\Cdot \ell_1)\cA_3^{h_2}(\pb_1,\ell_2)\cA_4^{h_1,h_2}(\pb_2,\ell_1,\ell_2)\, , 
\end{align}
with $\ell_2=-\ell_1+q$. The diagram above is $\cO(\mb_1^3\mb_2^2)$ and the $\cO(\mb_1^2\mb_2^3)$ contribution is found by swapping the particle labels $1$ and $2$. 
In principle, there are also hyperclassical contributions to the one-loop amplitude which come from gluing two copies of \eqref{eq: comptonheftm3}  together. However, these $\cO(\mb_1^3\mb_2^3)$ terms will cancel out in the observables we are interested in here. 

To compute the cut above we perform the graviton propagator sum using 
\begin{align}
	\label{eq: symGravStateSum}
	\sum_{h}\varepsilon^{\mu}_{-k} {\varepsilon}^{\nu}_{-k}{\varepsilon}^{\alpha}_{k} \varepsilon^{\beta}_{k}=\frac{1}{2}\Big( P^{\mu \alpha}P^{\nu \beta}+P^{\mu \beta}P^{\nu \alpha}\Big)-\frac{1}{D-2}P^{\mu\nu}P^{\alpha\beta}
	\ ,
\end{align}
where  
\begin{equation}
    P^{\mu\nu}\coloneqq \eta^{\mu\nu}-\frac{k^\mu t^\nu+k^\nu t^\mu}{k\Cdot t}\,,\quad k\Cdot\varepsilon_k=t\Cdot\varepsilon_k=0\,,
\end{equation}
  $k^\mu$ is the graviton momentum and $t^\mu$ is a null reference momentum transverse to $\varepsilon_k$. In practice, we can replace $P^{\mu\nu}\rightarrow\eta^{\mu\nu}$ at the beginning since the dependence on the reference momentum $t^\mu$  will drop out of the gauge invariant state sum.%
 \footnote{For the manipulation of expressions in this section, we used the Mathematica package DDimPackage \cite{DDims}.} However, the reference momentum dependence can be more subtle in the spinning case, as we will see later. 

After performing the state sum and simplifying the result, we lift the propagators $\ell_1, \ell_2$  off the cuts  (i.e.~we replace $\delta(\ell_i^2)\rightarrow i/\ell_i^2$), and use integration by parts (IBP) relations
to reduce the integrand.%
\footnote{We perform the IBP reductions in this paper using LiteRed \cite{Lee:2012cn,Lee:2013mka}.} We are then left with an amplitude that is proportional to the scalar cut triangle. In $D=4$ we have, 
\begin{equation}
    \cM_4=i\frac{3\pi}{32}\kappa^4\mb_1^3\mb_2^2(5\bar{\sigma}^2-1)I_3(\ub_1,q^2) + (1\leftrightarrow 2)\,,
\end{equation}
where 
\begin{equation}
\label{selfsimilar}
I_3(\ub_i,q^2)\coloneqq\int\!\frac{d^4\ell_1}{(2\pi)^4}\frac{\delta(\ub_i\Cdot\ell_1)}{\ell_1^2(\ell_1-q)^2}=\frac{1}{16 \pi }\frac{1}{ |q|}\,.
\end{equation}
Thus the final result for the one-loop classical four-point amplitude is 
\begin{equation}\label{eq: EH4point1loop}
    \cM_4=i\left( \frac{\kappa}{2}\right)^4
    \frac{3}{32}\mb_1^2\mb_2^2(\mb_1+\mb_2)\frac{5\bar{\sigma}^2-1}{|q|}\,,
\end{equation}
where we have defined $\bar{\sigma}\coloneq \ub_1\Cdot\ub_2$.

An important comment on barred vs.~unbarred variables is in order here. Since hyperclassical terms exist for the four-point amplitude above, the distinction between $\mb_i$ and $m_i$ is in principle important, since upon re-expanding in $m_i$ these hyperclassical terms could feed down into the classical amplitude above. However, in the elastic one-loop case we are considering, these feed-down terms only contribute to quantum corrections and do not affect the classical amplitude. This can be seen by noting that $\mb_i$ and $m_i$ differ by a term proportional to $q^2$ (see \eqref{mbardef}); furthermore the only other non-zero kinematic invariant that we can build out of the massive momenta is $\bar{\sigma}=\ub_1\Cdot\ub_2$, however due to four-point kinematics this also differs from $\sigma$ by a term proportional to $q^2$, as can be seen from \eqref{barredv}.
Thus we are free to remove the bars in \eqref{eq: EH4point1loop}, and a similar conclusion applies to the amplitudes in the presence of cubic corrections. 
 
\subsection{Introducing \texorpdfstring{$R^3$}{R3} corrections: the scalar case}
In this paper we are interested in the corrections to the classical scattering of two spinning black holes from the operators $I_1$ and $G_3$ (or equivalently $I_1$ and $I_2$). At leading order in their Wilson coefficients, these operators only enter through their corrections to the Compton amplitudes described in Section~\ref{sec: ClassicalComptonAmplitudes}. In particular, the three-point amplitudes involving two heavy particles and one graviton are left unchanged. In the heavy-mass expansion, the $R^3$ corrections to the Compton amplitudes begin at $\cO(\mb^2)$, not $\cO(\mb^3)$ as for the full Compton amplitudes Section~\ref{sec: ClassicalComptonAmplitudes}. This is due to a lack of massive propagators in the $R^3$ Compton amplitudes, which are responsible for the $\cO(\mb^3)$ terms in \eqref{eq: comptonheftm3}.

For the reasons above, the first corrections to the classical four-point massive amplitude due to $R^3$ interactions begin at one loop, not tree level. To compute these we will again first focus on the scalar case. The procedure is identical to the one detailed in the previous section: perform a cut in the $q^2$-channel, then glue heavy-mass expanded amplitudes together to obtain the classical piece, and finally, IBP reduce the result. The only difference is that now we use the $R^3$ corrections to the Compton amplitude at $\cO(\mb^2)$. Explicitly we have the $\cO(\mb_1^2 \mb_2^2)$ cut 
\begin{align}\label{eq: scalarR3FourPoint}
    &\begin{tikzpicture}[baseline={([yshift=-0.4ex]current bounding box.center)}]\tikzstyle{every node}=[font=\small]
            \node (b1) at (1,0) {};
            \node (b3) at (3,0) {};
            \node (u1) at (2,2.5) {};
            \draw[double, vector, line width = .75] (b1) -- (u1);
            \draw[double, vector, line width = .75] (b3) -- (u1);
		      \begin{feynman}
			\vertex (p1) {\(p_1\)};
			\vertex [above=2.5cm of p1](p2){$p_2$};
			\vertex [right=2cm of p2] (u1) [HV]{$R^3$};
			\vertex [right=2cm of u1] (p3){$p_{2'}$};
			\vertex [right=1.0cm of p1] (b1) [HV]{H};
			\vertex [right=1.0cm of b1] (b2) []{};
			\vertex [right=1.0cm of b2] (b3) [HV]{H};
			\vertex [right=1.0cm of b3](p4){$p_{1'}$};
			\vertex [above=1.25cm of p1] (cutL);
			\vertex [right=4.0cm of cutL] (cutR){};
			\vertex [right=1.0cm of b1] (cut1);
			\vertex [above=0.3cm of cut1] (cut1u);
			\vertex [below=0.3cm of cut1] (cut1b);
			\vertex [right=0.5cm of b2] (cut2);
			\vertex [above=0.3cm of cut2] (cut2u);
			\vertex [below=0.3cm of cut2] (cut2b);
			\diagram* {
			(p2) -- [thick] (u1) -- [thick] (p3),
			(b1)--[opacity=0,rmomentum=\(\ell_{1}\)](u1), (b3)-- [opacity=0,rmomentum'=\(\ell_{2}\)] (u1), (p1) -- [thick] (b1)-- [thick] (b3)-- [thick] (p4), (cutL)--[dashed, red,thick] (cutR), (cut1u)--[ red,thick] (cut1b),
			};
		\end{feynman}
	      \end{tikzpicture}\nn\\
       &=\frac{1}{2}\sum_{h_1,h_2}\int\frac{d^D\ell_1}{(2\pi)^D}\delta(\ell_1^2)\delta(\ell_2^2)\cA_3^{h_1}(\pb_1,\ell_1)\pi\delta(\pb_1\Cdot \ell_1)\cA_3^{h_2}(\pb_1,\ell_2)M_{4,R^3}^{h_1,h_2}(\pb_2,\ell_1,\ell_2)
\end{align}
where the $R^3$ correction to the classical Compton amplitude, $M_{4,R^3}$, is either \eqref{I1amplitude} or \eqref{I2amplitude} in the spinless limit. As before there is another $\cO(\mb_1^2\mb_2^3)$ cut obtained by swapping the particle labels 1 and 2. After performing the state sum using \eqref{eq: symGravStateSum} and then  IBP reducing the result, we obtain an integrand proportional to the same scalar triangle as before. For $I_1$ and $G_3$ corrections we have 
\begin{align}
\begin{split}
     \cM_{4,I_1}&=3i\pi\left(\frac{\kappa}{2}\right)^6\mb_1^3\mb_2^2(\bar{\sigma}^2-1) (q^2)^2 I_3(\ub_1,q^2) + (1\leftrightarrow 2)\\
     &=i\left(\frac{\kappa}{2}\right)^6\frac{3}{16}\mb_1^2\mb_2^2(\mb_1+\mb_2)|q|^3(\bar{\sigma}^2-1) \,,
      \end{split}
\end{align}
and 
\begin{align}
\begin{split}
    \cM_{4,G_3}&=-3i\pi\left(\frac{\kappa}{2}\right)^6\mb_1^3\mb_2^2 (q^2)^2 I_3(\ub_1,q^2) + (1\leftrightarrow 2)\\
    &=-i\left(\frac{\kappa}{2}\right)^6\frac{3}{16}\mb_1^2\mb_2^2(\mb_1+\mb_2)|q|^3\,.
      \end{split}
\end{align}
 As previously mentioned we can replace $\mb_i$ with $m_i$ and $\ub_i$ with $u_i$. In particular, in this case there are no hyperclassical $R^3$ corrections which could feed down to the classical terms above. After performing such a replacement, the above expressions match the results of \cite{Brandhuber:2019qpg}.

\subsection{Introducing \texorpdfstring{$R^3$}{R3} corrections: the spinning case}
To add spin to the one-loop classical four-point $R^3$ amplitude we must add spin dependence to the two Compton amplitudes that appear on either side of the $q^2$-cut. First, let us examine the expansion of spinning Compton amplitudes in the classical limit. Here we focus on the case of the Compton amplitude with two positive (or two negative) helicity gravitons, as this is the only contribution we will need for the $R^3$ computation. 

Ideally, we wish to start from an unexpanded finite-spin Compton amplitude and then take the classical limit $ s/m{\rightarrow}  |a|$, $s{\rightarrow}\infty$. The  spinning Compton amplitudes in EH gravity with two positive (or two negative) helicity gravitons can be found by demanding appropriate factorisation into three-point amplitudes~\cite{Aoude:2020onz,Johansson:2019dnu,Lazopoulos:2021mna},
\begin{align}
    &M_4(-\boldsymbol{p}^s,\boldsymbol{p'}^{s},k_1^{++},k_2^{++})= \frac{\langle \boldsymbol{p}\,\boldsymbol{p'} \rangle^{ 2 s}}{m^{2s}} M_4(-p,p',k_1^{++},k_2^{++})\,,\\
    &M_4(-\boldsymbol{p}^s,\boldsymbol{p'}^{s},k_1^{--},k_2^{--})= \frac{[\boldsymbol{p}\,\boldsymbol{p'} ]^{ 2 s}}{m^{2s}} M_4(-p,p',k_1^{--},k_2^{--})\,,
\end{align}
where the massive spinors 
$|\boldsymbol{p}^s\rangle$ and $ |\boldsymbol{p'}^{s}\rangle$ are defined in \eqref{defboldp} (with similar definitions for $|\boldsymbol{p}^s]$ and $ |\boldsymbol{p'}^{s}]$),
and $M_4(p,p',k_1^{\pm\pm},k_2^{\pm\pm})$ is the scalar Compton amplitude. We will only need the leading piece in the classical expansion of the above amplitude for our purposes. This will be the $\cO(\mb^3)$ piece of the scalar Compton amplitude multiplied by the large spin (and large mass) limit of the spinning prefactor above. Explicitly, in an analogous manner to \eqref{eq:CLmassivecut} with $l_i = -k_i$,  we have
\begin{align}\label{eq: comptonSpinningm3}
\begin{split}
    M_{4,\mb^3}(\pb,a,k_1^{\pm\pm},k_2^{\pm\pm})= e^{\pm(k_1+k_2)\Cdot a} \cA_3(\pb,k_1^{\pm\pm})\,\pi\delta(\pb\Cdot k_1)\,\cA_3(\pb,k_2^{\pm\pm})\,,
    \end{split}
    \end{align}
    with 
    \begin{align}
    \cA_{3}(\pb,k_i^{\pm\pm})\coloneqq -i\kappa (\pb\Cdot \varepsilon_{k_i}^\pm)^2\,.
    \end{align}
The spinning $R^3$ Compton amplitudes are $\cO(\mb^2)$ to the leading order in the classical limit. Additionally, as we mentioned in Section~\ref{sec: ClassicalComptonAmplitudes},  they vanish for $(+,-)$ and $(-,+)$ helicity configurations. 
As a result, to obtain a classical contribution to the four-point spinning one-loop amplitude we need to glue the $\cO(\mb^2)$ $R^3$ Compton amplitudes to the $\cO(\mb^3)$ equal-helicity amplitude in \eqref{eq: comptonSpinningm3}
\begin{align}\label{eq: R3SpinningCut}
\frac{e^{-q\cdot a_1}}{2}\times
    \begin{tikzpicture}[baseline={([yshift=-0.4ex]current bounding box.center)}]\tikzstyle{every node}=[font=\small]
            \node (b1) at (1,0) {};
            \node (b3) at (3,0) {};
            \node (u1) at (2,2.5) {};
            \draw[double, vector, line width = .75] (b1) -- (u1);
            \draw[double, vector, line width = .75] (b3) -- (u1);
		      \begin{feynman}
			\vertex (p1) {\(p_1\)};
			\vertex [above=2.5cm of p1](p2){$p_2$};
			\vertex [right=2cm of p2] (u1) [HV]{$R^3,a_2$};
                \vertex [below=1.05cm of u1] (hp3){};
                \vertex [left=0.2cm of hp3] (h3){$-$};
                \vertex [right=0.2cm of hp3] (h4){$-$};
			\vertex [right=2cm of u1] (p3){$p_{2'}$};
			\vertex [right=1.0cm of p1] (b1) [HV]{H};
                \vertex [right=0.5cm of b1] (hp1) {};
                \vertex [above=0.5cm of hp1] (h1) {+};
			\vertex [right=1.0cm of b1] (b2) []{};
			\vertex [right=1.0cm of b2] (b3) [HV]{H};
                \vertex [left=0.5cm of b3] (hp2) {};
                \vertex [above=0.5cm of hp2] (h2) {+};
			\vertex [right=1.0cm of b3](p4){$p_{1'}$};
			\vertex [above=1.25cm of p1] (cutL);
			\vertex [right=4.0cm of cutL] (cutR){};
			\vertex [right=1.0cm of b1] (cut1);
			\vertex [above=0.3cm of cut1] (cut1u);
			\vertex [below=0.3cm of cut1] (cut1b);
			\vertex [right=0.5cm of b2] (cut2);
			\vertex [above=0.3cm of cut2] (cut2u);
			\vertex [below=0.3cm of cut2] (cut2b);
			\diagram* {
			(p2) -- [thick] (u1) -- [thick] (p3),
			(b1)--[opacity=0](u1), (b3)-- [opacity=0] (u1), (p1) -- [thick] (b1)-- [thick] (b3)-- [thick] (p4), (cutL)--[dashed, red,thick] (cutR), (cut1u)--[ red,thick] (cut1b),
			};
		\end{feynman}
	      \end{tikzpicture}
      & +\frac{e^{+q\cdot a_1}}{2}\times
    \begin{tikzpicture}[baseline={([yshift=-0.4ex]current bounding box.center)}]\tikzstyle{every node}=[font=\small]
            \node (b1) at (1,0) {};
            \node (b3) at (3,0) {};
            \node (u1) at (2,2.5) {};
            \draw[double, vector, line width = .75] (b1) -- (u1);
            \draw[double, vector, line width = .75] (b3) -- (u1);
		      \begin{feynman}
			\vertex (p1) {\(p_1\)};
			\vertex [above=2.5cm of p1](p2){$p_2$};
			\vertex [right=2cm of p2] (u1) [HV]{$R^3,a_2$};
                \vertex [below=1.05cm of u1] (hp3){};
                \vertex [left=0.2cm of hp3] (h3){$+$};
                \vertex [right=0.2cm of hp3] (h4){$+$};
			\vertex [right=2cm of u1] (p3){$p_{2'}$};
			\vertex [right=1.0cm of p1] (b1) [HV]{H};
                \vertex [right=0.5cm of b1] (hp1) {};
                \vertex [above=0.5cm of hp1] (h1) {$-$};
			\vertex [right=1.0cm of b1] (b2) []{};
			\vertex [right=1.0cm of b2] (b3) [HV]{H};
                \vertex [left=0.5cm of b3] (hp2) {};
                \vertex [above=0.5cm of hp2] (h2) {$-$};
			\vertex [right=1.0cm of b3](p4){$p_{1'}$};
			\vertex [above=1.25cm of p1] (cutL);
			\vertex [right=4.0cm of cutL] (cutR){};
			\vertex [right=1.0cm of b1] (cut1);
			\vertex [above=0.3cm of cut1] (cut1u);
			\vertex [below=0.3cm of cut1] (cut1b);
			\vertex [right=0.5cm of b2] (cut2);
			\vertex [above=0.3cm of cut2] (cut2u);
			\vertex [below=0.3cm of cut2] (cut2b);
			\diagram* {
			(p2) -- [thick] (u1) -- [thick] (p3),
			(b1)--[opacity=0](u1), (b3)-- [opacity=0] (u1), (p1) -- [thick] (b1)-- [thick] (b3)-- [thick] (p4), (cutL)--[dashed, red,thick] (cutR), (cut1u)--[ red,thick] (cut1b),
			};
		\end{feynman}
	      \end{tikzpicture}\nn\\
       \nn\\
       &\mkern-18mu\mkern-18mu\mkern-18mu\mkern-18mu\mkern-18mu\mkern-18mu\mkern-18mu\coloneq e^{-q\Cdot a_1}\frac{1}{2} \cC(+,+) + e^{+q\Cdot a_1}\frac{1}{2} \cC(-,-)
\end{align}
where $\cC(\pm,\pm)$ denotes the cuts according to the helicity of graviton emitted from the massive $p_1$ line. 

Note that, unlike in the triple cut of Figure~\ref{fig:LS},  the two three-point amplitudes on the bottom line are indistinguishable since they are symmetric under swapping $\ell_1$ and $\ell_2$, thus we still need the factor of $1/2$ for summing over bosons.

We can rewrite this cut combination in terms of $\cosh(a_1\Cdot q)$ and $\sinh(a_1\Cdot q)$ as follows, 
\begin{equation} \label{eq: coshSinhCuts}
    \cosh(a_1\Cdot q) \frac{1}{2}\left(\cC(+,+)+\cC(-,-)\right)- \sinh(a_1\Cdot q)\frac{1}{2}\left(\cC(+,+)-\cC(-,-)\right)\,.
\end{equation}
To compute the cut contributions, we will rewrite the above expression in terms of symmetric, and anti-symmetric graviton state sums. To do this, we first note that $\cC(-,+)=\cC(+,-)=0$ due to the vanishing of the $R^3$ Compton amplitude for these helicity configurations. This allows us to rewrite the $\cosh(a_1\Cdot q)$ terms as 
\begin{align}\label{eq: cosha1Term}
  &\frac{1}{2}\left(\cC(+,+)+\cC(-,-)\right)=\frac{1}{2}\left(\cC(+,+)+\cC(+,-)+\cC(-,+)+\cC(-,-)\right)\\
    &=\frac{1}{2}\sum_{h_1,h_2}\int\frac{d^D\ell_1}{(2\pi)^D}\delta(\ell_1^2)\delta(\ell_2^2)\cA_3^{h_1}(\pb_1,\ell_1)\pi\delta(\pb_1\Cdot \ell_1)\cA_3^{h_2}(\pb_1,\ell_2)M_{4,R^3}^{h_1,h_2}(\pb_2,a_2,\ell_1,\ell_2)\,,\nn
\end{align}
and similarly for the $\sinh(a_1\Cdot q)$ terms
\begin{align}\label{eq: sincha1Term}
     &\frac{1}{2}\left(\cC(+,+)-\cC(-,-)\right)=\frac{1}{2}\left(\cC(+,+)-\cC(+,-)+\cC(-,+)-\cC(-,-)\right)\\
    &=\frac{1}{2}\sum_{h_1,h_2}\frac{h_2}{2}\int\frac{d^D\ell_1}{(2\pi)^D}\delta(\ell_1^2)\delta(\ell_2^2)\cA_3^{h_1}(\pb_1,\ell_1)\pi\delta(\pb_1\Cdot \ell_1)\cA_3^{h_2}(\pb_1,\ell_2)M_{4,R^3}^{h_1,h_2}(\pb_2,a_2,\ell_1,\ell_2)\, ,\nn
\end{align}
where as usual the sums over helicity take the values $h_i=\pm 2$. The $\cosh(a_1\Cdot q)$ term \eqref{eq: cosha1Term} contains only ``symmetric'' graviton state sums in $h_1$ and $h_2$. By symmetric state sum here we mean the standard completeness relation 
\begin{equation}
    \varepsilon_{-k}^{\mu+} \varepsilon_{-k}^{\nu +} \varepsilon_{k}^{\alpha-} \varepsilon_{k}^{\beta-} + \varepsilon_{-k}^{\mu-} \varepsilon_{-k}^{\nu-}\varepsilon_{k}^{\alpha+} \varepsilon_{k}^{\beta+} =\frac{1}{2}\Big( P^{\mu \alpha}P^{\nu \beta}+P^{\mu \beta}P^{\nu \alpha}\Big)-\frac{1}{2}P^{\mu\nu}P^{\alpha\beta}\,,
\end{equation}
which is \eqref{eq: symGravStateSum} in the specific case of $D=4$. However, the expression for the $\sinh(a_1\Cdot q)$ term \eqref{eq: sincha1Term} contains an antisymmetric state sum in $h_2$ which evaluates to
\begin{align}
    &\varepsilon_{-k}^{\mu+} \varepsilon_{-k}^{\nu +} \varepsilon_{k}^{\alpha-} \varepsilon_{k}^{\beta-} - \varepsilon_{-k}^{\mu-} \varepsilon_{-k}^{\nu-}\varepsilon_{k}^{\alpha+} \varepsilon_{k}^{\beta+} =\frac{1}{4}\bigg(P^{\alpha\mu } \tilde{P}^{\beta\nu }+P^{\beta\nu } \tilde{P}^{\alpha\mu }+P^{\beta\mu } \tilde{P}^{\alpha\nu }+P^{\alpha\nu } \tilde{P}^{\beta\mu }
    \bigg)\,,
\end{align}
where
\begin{equation}\label{eq: antiSymGravStateSum}
    \tilde{P}^{\mu\nu}\coloneqq i\frac{ \epsilon^{\mu\nu\rho\sigma}k_\rho t_\sigma}{k\Cdot t}\,.
\end{equation}
Performing these state sums, we find that the term proportional to $\sinh(a_1\Cdot q)$ in the sum \eqref{eq: coshSinhCuts} vanishes, and thus the cut \eqref{eq: R3SpinningCut} is proportional to $\cosh{(a_1\Cdot q)}$. %
After lifting off the cut condition we have a tensor triangle integral which we reduce using Passarino-Veltman reduction followed by IBP relations. This, once again, leaves us with a result proportional to the scalar triangle
\begin{align}
\label{eq:a4I1class}
     \cM_{4,I_1}
     &=i\left(\frac{\kappa}{2}\right)^6\frac{3}{16}\mb_1^3\mb_2^2|q|^3 \cosh(a_1\Cdot q)\bigg[ (\bar{\sigma}^2{-}1)\cosh(a_2\Cdot q)- i \bar{\sigma}\frac{\sinh(a_2\Cdot q)}{a_2\Cdot q} \epsilon(\ub_1 \ub_2 a_2 q)
 \bigg]\nn\\
 & + (1{\leftrightarrow}2)\,,\nn\\
\end{align}
and 
\begin{align}
\begin{split}
\label{eq:a4G3class}
    \cM_{4,G_3}
    &=-i\left(\frac{\kappa}{2}\right)^6\frac{3}{16}\mb_1^3\mb_2^2 |q|^3 \cosh(a_1\Cdot q)\cosh(a_2\Cdot q)+ (1\leftrightarrow 2)\,.
\end{split}
\end{align}
As before, we can remove the bars from the above expressions and by doing so we recover our earlier results \eqref{G3amp} and \eqref{triple-cut-result}.

\section{Parity-odd cubic deformations}
\label{sec:parity-odd}

So far we have considered only corrections to general relativity  due to parity-even structures that are cubic in the Riemann tensor, $I_1$ and $I_2$ (or $G_3)$.
It is easy to repeat this analysis for parity-odd versions, $\tilde{I}_1$ and $\tilde{I}_2$, which can for example arise from integrating out (chiral) matter coupled to gravity.
These interactions are obtained from their parity-even cousins by simply replacing one of the three Riemann tensors by its dual version
\begin{align}
\tilde R^{\mu\nu\alpha\beta} & \coloneq \frac{1}{2} \epsilon^{\mu\nu\rho\sigma} {R_{\rho\sigma}}^ {\alpha\beta} \ .
\end{align}
This will affect the Compton amplitudes that enter the unitarity cuts/leading singularity computations performed earlier.
First, we notice that we only need to understand how on-shell gravitons are affected. When we compute the parity-odd versions of the vertices \eqref{V1} and \eqref{V2} we produce terms where the Levi-Civita tensor acts on the off-shell graviton but we can always interpret this Levi-Civita tensor to act on one of the on-shell gravitons. For $\tilde{I}_1$ this is immediate, while for $\tilde{I}_2$ this requires some algebra.

In essence we only have to replace one of the on-shell gravitons:
\begin{align}
R_i^{\alpha\beta\mu\nu} = \frac{\kappa}{2} F_i^{\alpha\beta} F_i^{\mu\nu} & \rightarrow
\tilde{R}_i^{\alpha\beta\mu\nu} = \frac{\kappa}{2} \tilde{F}_i^{\alpha\beta} F_i^{\mu\nu} \ ,
\end{align}
where 
\begin{align}
\tilde F^{\alpha\beta} & = \frac{1}{2} \epsilon^{\alpha\beta\mu\nu} F_{\mu\nu} \ .
\end{align}
Since we are in four dimensions, the gravitons can carry two possible helicities $+$ or $-$, which correspond to anti-selfdual (ASD) and selfdual (SD) field strengths, respectively.
In the case of positive helicity (ASD), we have
\begin{align}
F^{(+)} = -\ast F^{(+)} = -i \tilde{F}^{(+)} \Rightarrow \tilde{F}^{(+)} = i F^{(+)} \ , 
\end{align}
where $\ast$ is the Hodge star operation.
In the case of negative  helicity (SD), we have
\begin{align}
F^{(-)} = \ast F^{(-)} = i \tilde{F}^{(-)} \Rightarrow \tilde{F}^{(-)} = -i F^{(-)} \ .
\end{align}
As in the parity-even case, both on-shell gravitons must have the same helicity otherwise the vertex/Compton amplitude vanishes. Hence the overall effect is that, for the case of positive helicity, we get an overall factor of $+i$; while for negative helicity we get an overall factor of $-i$ for the Compton amplitude. Note that it does not matter which on-shell graviton is replaced by its dual version since both must have the same helicity.

If we now sum in the cut/leading singularity over the two allowed helicity configurations
we get the overall factor
\begin{align}
(+i) e^{a_1 \Cdot q} + (-i) e^{-a_1 \Cdot q} =
2 i \sinh(a_1 \Cdot q) \ ,
\end{align}
instead of $2 \cosh(a_1 \Cdot q)$ in the case of parity-even interactions.

Summarising, in order to get the corrections from the parity-odd, cubic curvature terms we simply have to replace in \eqref{triple-cut-result} the factor $\cosh(a_1 \Cdot q)$ by
$i \sinh(a_1 \Cdot q)$. Doing so, we arrive at the results:
\begin{align}
\label{parity-odd-triple-cut-result}
\begin{split}
\cM_{\tilde{I}_1}^{\bigtriangleup}& = -\left(\frac{\kappa}{2}\right)^6
    \frac{3}{16} m_1^3 m_2^2 |q|^3  \sinh (a_1 \Cdot q)
    \left[ (\sigma^2-1) \cosh (a_2 \Cdot q) - i \sigma \frac{\sinh (a_2 \Cdot q)}{a_2 \Cdot q}
    \epsilon(u_1 u_2 a_2 q)\right] , \\
    \cM_{\tilde{G}_3}^{\bigtriangleup} & = \left(\frac{\kappa}{2}\right)^6
    \frac{3}{16} m_1^3 m_2^2 |q|^3 \, \sinh (a_1 \Cdot q)
     \cosh (a_2 \Cdot q) \, .
    \end{split}
\end{align}
As before, the complete amplitudes can be obtained using \eqref{completeamp} and  \eqref{completeamp2} (see also  \eqref{swapped}). Note however the additional sign that appears compared to the 
parity-even case,
\begin{align}
\label{parity-odd-triple-cut-result-bis}
\begin{split}
\cM_{\tilde{I}_1}^{\bigtriangledown}& = \left(\frac{\kappa}{2}\right)^6
    \frac{3}{16} m_1^2 m_2^3 |q|^3  \sinh (a_2 \Cdot q)
    \left[ (\sigma^2-1) \cosh (a_1 \Cdot q) - i \sigma \frac{\sinh (a_1 \Cdot q)}{a_1 \Cdot q}
    \epsilon(u_1 u_2 a_1 q)\right] , \\
    \cM_{\tilde{G}_3}^{\bigtriangledown} & = -  \left(\frac{\kappa}{2}\right)^6
    \frac{3}{16} m_1^2 m_2^3 |q|^3 \, \sinh (a_2 \Cdot q)
     \cosh (a_1 \Cdot q) \, .
    \end{split}
\end{align}

\section{Observables from KMOC}
\label{sec:Obs-from-KMOC}

We begin with a lightning review of the KMOC approach to classical observables, following \cite{Kosower:2018adc, Maybee:2019jus} 
 (see also 
\cite{Guevara:2019fsj, Gatica:2023iws, Luna:2023uwd} for some recent applications in the presence of spin). 
For any observable $\mathbb{O}$, we want to compute the change  
\begin{align}
    \langle  \Delta \mathbb{O}  \rangle_{\psi} = \mbox{}_{\rm out}\langle \psi | \mathbb{O} | \psi \rangle_{\rm out}- \mbox{}_{\rm in}\langle \psi | \mathbb{O} | \psi \rangle_{\rm in} = \mbox{}_{\rm in}\langle \psi | S^\dagger \mathbb{O} S| \psi \rangle_{\rm in}- \mbox{}_{\rm in}\langle \psi | \mathbb{O} | \psi \rangle_{\rm in}\ . 
    \end{align}
Using  $S = \mathbbm{1} + i T$ and unitarity, one arrives at \cite{Kosower:2018adc}
\begin{align}
\label{deltaO}
    \langle  \Delta \mathbb{O}  \rangle_{\psi}  = i \ \mbox{}_{\rm in}\langle \psi | [ \mathbb{O},  T ] | \psi \rangle_{\rm in}+  \mbox{}_{\rm in}\langle \psi | T^\dagger [ \mathbb{O} , T]| \psi \rangle_{\rm in}\ . 
    \end{align}
We represent the initial state of the two heavy objects  as  
\begin{align}
	\label{psi}
	|\psi\rangle_{\rm in} \coloneqq \int\!d\Phi(p_1) d\Phi(p_2) e^{i (p_1 \Cdot b_1 + p_2 \Cdot b_2)} \phi(p_1) \phi(p_2) \xi_{1, i_1} \xi_{2, i_2} |p_1, i_1,   p_2, i_2\rangle_{\rm in}
	\ ,
\end{align}
with  the wavefunctions $\phi(p_1)$ and $\phi(p_2)$ being  peaked around the classical values of the momenta. We also define  
\begin{align}
\label{dphik}
	\begin{split}
		d\Phi(p) & \coloneqq \frac{d^Dp}{(2\pi)^{D{-}1} }\delta^{(+)} (p^2 - m^2)\, , \qquad 
		|p,i\rangle  \coloneqq a^\dagger_i (\vec{p})  |0\rangle\, , \end{split}
\end{align}
with $[a_i (\vec{p}) , a^\dagger_j (\vec{p}^{\, \prime})] = (2\pi)^{D{-}1} (2 E_p) \delta_{ij} \delta^{(D{-}1)}(\vec{p} - \vec{p}^{\, \prime})$, with the  normalisations 
\begin{align}
    \int\!d\Phi (p) \, |\phi(p)|^2 = \sum_i |\xi_i|^2 = 1\, . 
\end{align}
The state $|p_1, i_1,   p_2, i_2\rangle_{\rm in}$ defines two incoming particles with momenta $p_1$ and $p_2$, and spins $s_1$ and $s_2$ respectively. The indices $i_1$, $i_2$ encode the spin degrees of freedom of the two particles, namely they are the little group indices in the spin-$s_1$ and spin-$s_2$ representation, such that $i_k = 1,\dots,2s_k+1$. The little group vectors $\xi_{1,i_1}$ and $\xi_{2,i_2}$ define a choice of polarisation for the two particles. They are related to the $SU(2)$ polarisation variables $z^{(p)}_a$ introduced in \eqref{defboldp} by a change of basis: namely, the symmetrised tensor product of $2s$ spinors, parameterised by the spinor polarisation variables $z^{(p)}_a$ with $a = 1,2$, yields the $2s+1$ components encoded in the variables $\xi_i$.

Any expectation value of an operator $\mathbb{O}$ can then be written as  
\begin{align}
\begin{split}
\label{Opsi}
    \langle \mathbb{O}\rangle_\psi &= \int\! d\Phi (p_1) d\Phi (p_2)  \phi (p_1) \phi (p_2)    \int\!\frac{d^D q_1}{(2\pi)^{D{-}1}} \frac{d^D q_2}{(2\pi)^{D{-}1}} \delta( 2\bar{p}_1\Cdot q_1 ) \delta( 2\bar{p}_2\Cdot q_2 ) e^{i(q_1\Cdot b_1 + q_2 \Cdot b_2)}\\
		&\phi^\ast (p_1 {-}  q_1)\phi^\ast (p_2 {-}  q_2)\  \bar{\xi}_{1,i^\prime_1}\bar{\xi}_{2,i^\prime_2}  
		\langle p_1^\prime,i_1',   p_2^\prime, i_2'   |  \mathbb{O} | p_1, i_1,   p_2 , i_2 \rangle  \xi_{1, i_1} \xi_{2, i_2} 
  \\ & \!\!\to 
  \int\! d\Phi (p_1) d\Phi (p_2)  |\phi (p_1)|^2 |\phi (p_2)|^2
  \int\!\frac{d^D q_1}{(2\pi)^{D{-}1}} \frac{d^D q_2}{(2\pi)^{D{-}1}} \delta( 2\bar{p}_1\Cdot q_1 ) \delta( 2\bar{p}_2\Cdot q_2 ) e^{i(q_1\Cdot b_1 + q_2 \Cdot b_2)}
  \\ &
    \bar{\xi}_{1,i^\prime_1}\bar{\xi}_{2,i^\prime_2}\langle p_1^\prime,i_1',   p_2^\prime, i_2'   |  \mathbb{O} | p_1, i_1,   p_2 , i_2 \rangle\xi_{1,i_1}\xi_{2,i_2} 
  \, ,
	\end{split}
\end{align}
where, 
as usual, in the last step  
we have approximated  $\phi (p_i {-} q_i) {\to} \phi (p_i)$, with $i=1,2$. 
The barred variables are defined~as 
\begin{align}
	\label{barredv-bis}
	\begin{split}
		p_1 &= \bar{p}_1 + \frac{q_1}{2}\, , \qquad p_1^\prime = \bar{p}_1 - \frac{q_1}{2}\, , \\
		p_2 &= \bar{p}_2 +  \frac{q_2}{2}\, , \qquad p_2^\prime = \bar{p}_2 - \frac{q_2}{2}\, , 
	\end{split}
\end{align}
and satisfy  \begin{align}
	\label{eq: HEFTfame-bis}
	\bar{p}_1\Cdot q_1 =\bar{p}_2\Cdot q_2 =0 \, . 
\end{align}
In the elastic case, that we are studying, we can set 
$q_2 = -  q_1\coloneq q$.

Note that computing classical observables requires a change of variables encoding the spin degrees of freedom~\cite{Maybee:2019jus}. Namely, for each incoming state of the form
\begin{equation}
    |\psi\rangle = \int\!d\Phi(p) e^{i p \Cdot b} \phi(p) \xi_{i} |p, i\rangle ,
\end{equation}
we consider the mass-rescaled Pauli-Lubanski vector $\mathbb{S}^\mu$,
\begin{equation}
    \label{eq: PauliLubanskiDef}
    \mathbb{S}^\mu = \frac{1}{2 m} \epsilon^{\mu\nu\rho\sigma} \, \mathbb{P}_\nu \, \mathbb{M}_{\rho\sigma} ,
\end{equation}
where $\mathbb{M}_{\rho\sigma}$ are Lorentz generators, and compute its expectation value,
\begin{equation}
    \bra{\psi} \mathbb{S}^\mu \ket{\psi} = \int\!d\Phi(p)  | \phi (p) |^2 \,  
    \bar{\xi}_{i} S^\mu_{ij}(p) \xi_{j} ,
\end{equation}
where we used $\bra{p',j} \mathbb{S}^\mu \ket{p,i} = (2\pi)^{D{-}1}   \delta^{(D{-}1)}(\vec{p} - \vec{p}^{\, \prime})\, (2 E_p)\, S^\mu_{i j}(p)$.  
The vector 
\begin{align}
\label{eq:spinvecKMOC}
    S^\mu(p) \coloneqq \bar{\xi}_{i} S^\mu_{ij}(p) \xi_{j}
    \end{align} is the angular momentum vector of the classical particle, 
    and indeed it  corresponds to the spin vector of a Kerr black hole, as defined in \eqref{def:spinvector}.
    It is also convenient to use the ring radius vector $a^\mu(p) \coloneqq S^\mu (p)/m$ already introduced in Section~\ref{sec: ClassicalComptonAmplitudes}.

The amplitude $\xi_{1,i_1}\xi_{2,i_2}\bar{\xi}_{1,i^\prime_1}\bar{\xi}_{2,i^\prime_2}\langle p_1^\prime,i_1',   p_2^\prime, i_2'   |  \mathbb{O} | p_1, i_1,   p_2 , i_2 \rangle$ can be written in terms of massive polarisation tensors 
\begin{align}
\bep_p^{\mu_1 \dots \mu_s} \coloneq \xi_{p,i} \, \ep_{p,i}^{\mu_1 \dots \mu_s}\, , 
\end{align}
where the bold notation denotes the contraction with the little group variables $\xi_{p,i}$. The spin vector $S^\mu (p)$ in \eqref{eq:spinvecKMOC} can be written in terms of such polarisation tensors,
\begin{align}
\label{eq:spinvecS-first}
\begin{split}
S^\mu(p) 
=& \bra{\psi} \mathbb{S}^\mu \ket{\psi} 
= (\bar{\bep}_p)_{\alpha(s)} {(\mathbb{S}^\mu)^{\alpha(s)}}_{\beta(s)} (\bep_p)^{\beta(s)} \\
=& \frac{1}{2m} \epsilon^{\mu\nu\rho\sigma} p_\nu (\bar{\bep}_p)_{\alpha(s)} {(\mathbb{M}_{\rho\sigma})^{\alpha(s)}}_{\beta(s)} (\bep_p)^{\beta(s)}
,
\end{split}
\end{align}
where $\alpha(s) \coloneq \alpha_1 \dots \alpha_s$ and ${(\mathbb{M}_{\rho\sigma})^{\alpha(s)}}_{\beta(s)} =  i \, 2s \, {\delta^{(\alpha_1}}_{[\rho} \eta_{\sigma](\beta_1} \delta^{\alpha_2}_{\beta_2} \dots \delta^{\alpha_s)}_{\beta_s)}$, and we have omitted the integration $\int\!d\Phi(p)  | \phi (p) |^2$, as is customary when using  sharply peaked wavepackets $\phi(p)$. 
For convenience, we write the polarisation tensor as a product of vectors, 
\begin{align}
\bep_{p}^{\mu_1 \dots \mu_s} \to  \bep_{p}^{\mu_1}\dots \bep_{p}^{\mu_s}\, ,  
\end{align}
and hence, for a spin-$s$ particle we have%
\footnote{See e.g.~\cite{Cangemi:2022abk}.}
\begin{equation}
\label{eq:spinvecS}
S^\mu(p) = \frac{i s}{m} \, (\bar{\bep}_p\Cdot\bep_p)^{s-1}\epsilon^{\mu\nu\rho\sigma} p_\nu \bar{\bep}_{p\rho} \bep_{p\sigma} .
\end{equation}
In order to recast the amplitude in terms of such variables, one needs to write the polarisation vectors $\bep_{p_i'}^\mu$ as a boost acting of $\bep_{p_i}^\mu$, for instance~\cite{Cangemi:2023ysz}
\begin{equation}
    \bep_{p_1^\prime}^\mu = -\bar{\bep}_{p_1}^\mu - \frac{\bar{\bep}_{p_1} \Cdot q}{2 m_1^2 \sqrt{1-\frac{q^2}{4m_1^2}}} \Big(p_1^\mu + \frac{q^\mu}{2}  \Big) ,
\end{equation}
where $q^\mu = p_1'-p_1$.
This requires the identification 
\begin{align}
    \xi_{1',i} = \bar{\xi}_{1,i}\, , 
    \end{align}
    effectively an alignment of the polarisation of the incoming and outgoing quantum states describing a single classical object. Interestingly, this is automatic in the KMOC framework, as can be seen in \eqref{Opsi}.
    An analogous relation can be derived to express $\bep_{p_{2'}}^\mu$ in terms of $\bep_{p_2}^\mu$. Once the amplitude is given in terms of $\bep_{p_1}^\mu$ and $\bep_{p_2}^\mu$ only, relations such as \eqref{eq:spinvecS} can be employed to rewrite it as a function of the spin vectors $S_{1}^\mu$ and $S_{2}^\mu$.

One more step required to recover classical physics is  the limit $s_i {\to} \infty$ \cite{Guevara:2018wpp,Chung:2018kqs}. The angular momentum $J$ for a spin-$s$ quantum particle is of order $J {\sim}  \hbar\,  s$, therefore if we take the $\hbar \to 0$ limit but wish to keep $J$ finite, we need $s {\sim} 1/\hbar \to \infty$. From \eqref{eq:spinvecS}, we can see that this implies
\begin{equation}
    S^\mu(p) \sim \frac{1}{\hbar} , 
\end{equation}
which we have used in the previous sections.

\subsection{The momentum kick}
As a  first  example, we consider the momentum operator $\mathbb{P}_1$ of  particle $1$.  From \eqref{deltaO} we have  \cite{Kosower:2018adc}
\begin{align}
    \langle  \Delta \mathbb{P}_1^\mu  \rangle_{\psi} = \langle  \Delta \mathbb{P}_1^\mu  \rangle_{\psi}^{(1)}  + \langle  \Delta \mathbb{P}_1^\mu  \rangle_{\psi}^{(2)}
    \ ,  
    \end{align}
with
\begin{align}
\begin{split}
     \langle  \Delta \mathbb{P}_1^\mu  \rangle_{\psi}^{(1)}  = -
     \int\!\frac{d^D q_1}{(2\pi)^{D{-}1}} &\frac{d^D q_2}{(2\pi)^{D{-}1}} \delta( 2\bar{p}_1\Cdot q_1 ) \delta( 2\bar{p}_2\Cdot q_2 ) e^{i(q_1\Cdot b_1 + q_2 \Cdot b_2)}\, 
		q_1^\mu \, \\
&\times\bar{\xi}_{1,i^\prime_1}\bar{\xi}_{2,i^\prime_2}\langle p_1^\prime,i_1', p_2^\prime, i_2'|  i T | p_1, i_1, p_2 , i_2 \rangle \xi_{1,i_1}\xi_{2,i_2}
    \ ,  
    \end{split}
    \end{align}
   and 
\begin{align}
     &\langle  \Delta \mathbb{P}_1^\mu  \rangle_{\psi}^{(2)} = 
     \sum_X \int\!d\Phi(r_1)d\Phi(r_2)d\Phi(X) \int\!\frac{d^D q_1}{(2\pi)^{D{-}1}} \frac{d^D q_2}{(2\pi)^{D{-}1}} \delta( 2\bar{p}_1\Cdot q_1 ) \delta( 2\bar{p}_2\Cdot q_2 ) e^{i(q_1\Cdot b_1 + q_2 \Cdot b_2)}\,\notag \\
     &\times(r_1 {-} p_1)^\mu \bar{\xi}_{1,i^\prime_1}\bar{\xi}_{2,i^\prime_2} \, 
\langle p_1^\prime,i_1', p_2^\prime, i_2'|  T^\dagger |r_1, j_1, r_2, j_2,X \rangle  \langle r_1, j_1, r_2, j_2,X|  T | p_1, i_1, p_2 , i_2 \rangle  \xi_{1,i_1}\xi_{2,i_2} \notag
    \ ,   
\end{align}
or equivalently%
\footnote{Note that in our conventions  $\cM$ denotes a matrix element of~$iT$.}
\begin{align}
\label{dp1}
\begin{split}
     \langle  \Delta \mathbb{P}_1^\mu  \rangle_{\psi}^{(1)}  = \int\!d\mu^{(D)} \, 
      e^{iq\Cdot b}\, 
		q^\mu \, \bar{\xi}_{1,i^\prime_1}\bar{\xi}_{2,i^\prime_2} \cM_{i^\prime_1 i^\prime_2 i_1 i_2}(p_1, p_2\to   p_1^\prime, p_2^\prime) \xi_{1,i_1}\xi_{2,i_2}
    \ ,  
    \end{split}
    \end{align}
    \begin{align}
    \label{dp2}
    \begin{split}
     &\langle  \Delta \mathbb{P}_1^\mu  \rangle_{\psi}^{(2)}  = \sum_X \int\!d\Phi(r_1)d \Phi(r_2)d \Phi(X)
     \int\!d\mu^{(D)} \, 
      e^{iq\Cdot b}\, (2\pi)^D \delta^{(D)} (p_1 + p_2 - r_1 - r_2 - r_X)\, 
      \\ &\qquad
		(r_1 {-} p_1)^\mu  \, \bar{\xi}_{1,i^\prime_1}\bar{\xi}_{2,i^\prime_2}\cM^\ast_{i^\prime_1 i^\prime_2 j_1 j_2} (  p_1^\prime, p_2^\prime{\to} r_1, r_2, X)\cM_{j_1 j_2 i_1 i_2}( p_1, p_2{\to}   r_1, r_2, X)  \xi_{1,i_1}\xi_{2,i_2}
    \ .  
    \end{split}
    \end{align}
In these relations 
we 
have omitted the ubiquitous $\int\! d\Phi (p_1) d\Phi (p_2)  |\phi (p_1)|^2 |\phi (p_2)|^2$ integration, and   
have introduced the measure
\begin{align}
\begin{split}
\label{measureinnn}
	d\mu^{(D)} 
\coloneqq 
 \frac{\!\!\!\!d^Dq}{(2\pi)^{D-2}} 
	\,   \delta(2  {\pb}_1\Cdot q ) \delta(2  {\pb}_2\Cdot q )
 \, ,
 \end{split}
\end{align}
where $b\coloneq b_2 {-} b_1$, and we recall that $q=p_1^\prime- p_1$.

Finally, we note that the second KMOC term $\langle  \Delta \mathbb{P}_1^\mu  \rangle_{\psi}^{(2)}$ in \eqref{dp2} 
contains the product of two amplitudes, which will necessarily be tree amplitudes at our leading order.
Since  there are  no tree-level contributions with an insertion of an $R^3$ vertex to two-to-two scattering,  we conclude that  at leading order, the  momentum kick in the presence of cubic deformations is fully captured by the term 
$\langle  \Delta \mathbb{P}_1^\mu  \rangle_{\psi}^{(1)}$ in \eqref{dp1}.

\subsection{The spin kick}

Similarly to the momentum kick, we can obtain the spin kick from KMOC, following \cite{Maybee:2019jus}. The observable of interest here is $\mathbb{O} = \mathbb{S}_1^\mu$, the mass-rescaled Pauli-Lubanski vector from \eqref{eq: PauliLubanskiDef} of particle $1$.

Starting from \eqref{deltaO} and focusing on the first term, we get 
\begin{align}
\label{spin-kick}
    \langle  \Delta \mathbb{S}_1^\mu  \rangle_{\psi}^{(1)}& = i\int\! \frac{d^D q_1}{(2\pi)^{D{-}1}} \frac{d^D q_2}{(2\pi)^{D{-}1}} \delta( 2\bar{p}_1\Cdot q_1 ) \delta( 2\bar{p}_2\Cdot q_2 ) e^{i(q_1\Cdot b_1 + q_2 \Cdot b_2)}\notag\\
    &\times
\bar{\xi}_{1,i^\prime_1}\bar{\xi}_{2,i^\prime_2}
    \ 
    \langle p_1^\prime,i_1', p_2^\prime, i_2'|  \mathbb{S}^\mu_1 T - T \mathbb{S}^\mu_1  | p_1, i_1, p_2 , i_2 \rangle\xi_{1,i_1}\xi_{2,i_2}\, .
\end{align}
Inserting a complete set of states,
\begin{equation}
    \mathbbm{1} = \sum_{j_1,j_2} \int\! d\Phi(r_1)d\Phi(r_2) |r_1, j_1, r_2, j_2 \rangle \langle r_1, j_1, r_2, j_2|\, ,
\end{equation}
we can write \eqref{spin-kick} in terms of the matrix elements of $\mathbb{S}_1^\mu$ and $iT$, so that
\begin{align}
    \langle  \Delta \mathbb{S}_1^\mu  \rangle_{\psi}^{(1)} =\int\! &\frac{d^D q_1}{(2\pi)^{D{-}1}} \frac{d^D q_2}{(2\pi)^{D{-}1}} \delta( 2\bar{p}_1\Cdot q_1 ) \delta( 2\bar{p}_2\Cdot q_2 ) e^{i(q_1\Cdot b_1 + q_2 \Cdot b_2)} (2\pi)^D \delta^{(D)}(q_1+q_2)  \notag\\
    & \bar{\xi}_{1,i^\prime_1}\bar{\xi}_{2,i^\prime_2} \left(S^\mu_{1\, i_1^\prime j_1} (p^\prime_1) \cM_{j_1 i_2^\prime i_1 i_2}-\cM_{i_1^\prime i_2^\prime j_1 i_2} S^\mu_{1\, j_1 i_1} (p_1)\right)
\xi_{1,i_1}\xi_{2,i_2}\, , \label{eq: spinKickIndices}
\end{align}
where $\cM = \cM(p_1, p_2 \to p^\prime_1, p^\prime_2)$. The delta function enforces $q_2=-q_1\coloneq q$, and thus $p^\prime_1 = p_1 + q$, which is related by an infinitesimal Lorentz transformation to $p_1$, meaning that the spin
\begin{equation}
    S^\mu_{1\, ij} (p^\prime_1) = S^\mu_{1\, ij} (p_1 + q) = S^\mu_{1\, ij} (p_1) -  \frac{q\Cdot S_{1\, ij}(p_1)}{m_1^2} p^\mu_1 \,  + \cO(\hbar^2)\, .
\end{equation}
Similarly to the case of the momentum kick, 
at leading order only  $\langle  \Delta \mathbb{S}_1^\mu  \rangle_{\psi}^{(1)}$ contributes, since $\langle  \Delta \mathbb{S}_1^\mu  \rangle_{\psi}^{(2)}$ contains 
the product of two tree-level two-to-two amplitudes, and there is no tree-level contribution to this process from a cubic vertex. 

Summarising, the leading contribution to the spin kick becomes
\begin{equation}
    \langle  \Delta \mathbb{S}_1^\mu  \rangle_{\psi}^{(1)} =\int\! d\mu^{(D)} e^{iq\Cdot b} \left( \left[S_1^\mu(p_1),\cM\right]-\frac{q\Cdot S_1(p_1)}{m_1^2} p^\mu_1 \, \cM \right)\, .
\end{equation}
Here, the little group indices should be understood to be arranged as in \eqref{eq: spinKickIndices}. This cleanly decomposes the spin kick into a longitudinal and transverse part (by virtue of $p_1 \Cdot S_1(p_1)=0$). We can recast this in terms of derivatives of the Fourier-transformed amplitude as was done in \cite{Guevara:2019fsj} to give 
\begin{equation}
\label{spingchange1}
    \langle  \Delta \mathbb{S}_1^\mu  \rangle_{\psi}^{(1)} = \frac{i}{m_1} \left(p^\mu_1 a^\nu_1 \frac{\partial}{\partial b^\nu}+\epsilon^{\mu\nu\rho\sigma}p_{1\nu} a_{1\rho}\frac{\partial}{\partial a_1^\sigma}\right) \widetilde{\cM}(b)\, , 
\end{equation}
where 
\begin{align}
    \widetilde{\cM}(b)\coloneq \int\!\frac{d^Dq}{(2\pi)^{D-2}}
\delta(2  {\pb}_1\Cdot q ) \delta(2  {\pb}_2\Cdot q ) e^{iq\Cdot b} \cM\, , 
\end{align}
is the amplitude in impact parameter space. 
This takes advantage of the fact that under the integral we can make the replacement $i q_\mu \mapsto \partial/\partial b^\mu$, and that
$\left[A, f(B)\right] = \left[A,B\right]\frac{\partial f}{\partial B}$, with $A$ and $B$ being $S_1^\mu$, $S_1^\nu$, respectively and $f(B)$ the amplitude $\cM$. 
Finally, we also comment that, because of the absence of a hyperclassical contribution to $\cM$, we can safely remove the bars in $\pb_{1,2}$.

\subsection{Fourier transforms to impact parameter space}

First we recall that
\begin{align}
\int\!\frac{d^Dq}{(2\pi)^{D-2}}
\delta(2  {p}_1\Cdot q ) \delta(2  {p}_2\Cdot q ) e^{iq\Cdot b} (\cdots) = 
\frac{1}{4 m_1 m_2 \sqrt{\sigma^2-1}}\int\!\frac{d^{D-2}q}{(2\pi)^{D-2}}e^{- i \vec{q}\Cdot \vec{b}} (\cdots) 
\ , 
\end{align}
where  $\sigma=p_1\Cdot p_2/ (m_1 m_2)$. 
Note that the momentum transfer $q^\mu$ is orthogonal to the scattering plane, defined by the incoming momenta $p_1$ and $p_2$, hence the Fourier transform is a two-dimensional integral.
It is convenient to work in the centre of mass frame of the two bodies, where $q^\mu = (0,\vec{q})$. In order to extract classical observables, as discussed in the previous sections, we need to compute the Fourier transform $\widetilde{\cM}(b)$ of the classical amplitudes given in \eqref{eq:a4I1class}, \eqref{eq:a4G3class} and \eqref{parity-odd-triple-cut-result}. This requires computing the following Fourier transforms,
\begin{align}\label{eq: SinchCoshIntegrals2}
\mathcal{F}_1 \coloneq &\int\!\frac{d^{D-2}q}{(2\pi)^{D-2}}e^{- i \vec{q}\Cdot \vec{b}} \, |\vec{q}\, |^3 \cosh{\vec{a}_1\Cdot \vec{q}} \ \cosh{\vec{a}_2\Cdot \vec{q}} , \\
\tilde{\mathcal{F}}_1 \coloneq &\int\!\frac{d^{D-2}q}{(2\pi)^{D-2}}e^{- i \vec{q}\Cdot \vec{b}} \, |\vec{q}\, |^3 \sinh{\vec{a}_1\Cdot \vec{q}} \ \cosh{\vec{a}_2\Cdot \vec{q}} , \\
\label{I2intdef}
\mathcal{F}_2 \coloneq    &\int\!\frac{d^{D-2}q}{(2\pi)^{D-2}}e^{- i \vec{q}\Cdot \vec{b}}\,  |\vec{q}\, |^3 (\vec{\epsilon}_2 \Cdot \vec{q}) \cosh{\vec{a}_1\Cdot \vec{q}} \  \frac{\sinh{\vec{a}_2\Cdot \vec{q}}}{\vec{a}_2\Cdot \vec{q}} , 
\\
\label{I2tildeintdef}
\tilde{\mathcal{F}}_2 \coloneq    &\int\!\frac{d^{D-2}q}{(2\pi)^{D-2}}e^{- i \vec{q}\Cdot \vec{b}}\,  |\vec{q}\, |^3 (\vec{\epsilon}_2 \Cdot \vec{q}) \sinh{\vec{a}_1\Cdot \vec{q}} \ \frac{\sinh{\vec{a}_2\Cdot \vec{q}}}{\vec{a}_2\Cdot \vec{q}} \, , 
\end{align}
where $\vec{\epsilon}_2 \Cdot \vec{q} \coloneq \epsilon(u_1,u_2,a_2,q)$.
In order to do so, we can rewrite the hyperbolic functions as operators in impact parameter space,
\begin{align}
\begin{split}
\label{eq:sinhcoshoperators}
    \cosh{\vec{a}_i\Cdot \vec{q}} = \frac{1}{2} (e^{\vec{a}_i\Cdot \vec{q}}+e^{ -\vec{a}_i\Cdot \vec{q}}) \to  \frac{1}{2} (e^{i \vec{a}_i\Cdot \vec{\nabla}_b}+e^{-i \vec{a}_i\Cdot \vec{\nabla}_b}) , \\
    \sinh{\vec{a}_i\Cdot \vec{q}} = \frac{1}{2} (e^{\vec{a}_i\Cdot \vec{q}}-e^{ -\vec{a}_i\Cdot \vec{q}}) \to  \frac{1}{2} (e^{i \vec{a}_i\Cdot \vec{\nabla}_b}-e^{-i \vec{a}_i\Cdot \vec{\nabla}_b}) .
\end{split}
\end{align}
This reduces the problem to the evaluation of the following integrals,
\begin{align}
\begin{split}
\label{ccc}
  &  \int\!\frac{d^{D-2}q}{(2\pi)^{D-2}}e^{- i \vec{q}\Cdot \vec{b}} \, |\vec{q}\,|^3  , 
   \\
   & \int\!\frac{d^{D-2}q}{(2\pi)^{D-2}}e^{- i \vec{q}\Cdot \vec{b}} \, |\vec{q}\, |^3 \frac{1}{\vec{a}_2\Cdot \vec{q}} \, , 
\end{split}
\end{align}
and to include the additional factor of $\vec{\epsilon}_2 \Cdot\vec{q}$ from \eqref{I2intdef} and \eqref{I2tildeintdef} we simply express $\vec{q}$ as a derivative in $\vec{b}$. 
The first integral in \eqref{ccc} is a special case of a known result,%
\footnote{See e.g. (A.3) in \cite{Brandhuber:2019qpg}.}
\begin{equation}
\label{FT-simple}
    \int\!\frac{d^{d}q}{(2\pi)^{d}}e^{- i \vec{q}\Cdot \vec{b}} |\vec{q}\, |^\alpha = \left(\frac{2}{|\vec{b}|}\right)^{d+\alpha} \frac{\Gamma\left(\frac{d+\alpha}{2}\right)}{(4\pi)^{d/2} \Gamma\left(-\frac{\alpha}{2}\right)} \xrightarrow[d = 2]{\alpha = 3} \frac{9}{2\pi |\vec{b}|^5} .
\end{equation}
To evaluate the second, we use  $\vec{a}_2 = i \vec{\tilde{a}}_2$  and obtain

\begin{align}
\label{eq:schwingertrick}
    \int\!\frac{d^{D-2}q}{(2\pi)^{D-2}}e^{- i \vec{q}\Cdot \vec{b}} |\vec{q}\, |^3 \frac{1}{i\vec{\tilde{a}}_2\Cdot \vec{q} + \epsilon} &=
    \int_{0}^\infty\!dt\,  \int\!\frac{d^{D-2}q}{(2\pi)^{D-2}}e^{- i \vec{q}\Cdot (\vec{b} + t\vec{\tilde{a}}_2) - \epsilon t} |\vec{q}\, |^3 \nn\\
    &= \frac{9}{2\pi} \int_{0}^{\infty} dt\,  e^{-\epsilon t} \frac{1}{[(\vec{b}+ t\vec{\tilde{a}}_2)^2]^{5/2}} ,
\end{align}
where we have introduced a regulator $\epsilon$, to be taken to zero at a later stage, to ensure convergence of the integrals.

It is now convenient to act on \eqref{eq:schwingertrick} with the operator $\sinh{i \vec{a}_2 \Cdot \vec{\nabla}_b} = -\sinh{\vec{\tilde{a}}_2 \Cdot \vec{\nabla}_b}$, as defined in \eqref{eq:sinhcoshoperators}, to obtain
\begin{align}
\begin{split}
\mathcal{J}& \coloneq \int\!\frac{d^{D-2}q}{(2\pi)^{D-2}}e^{- i \vec{q}\Cdot \vec{b}}\,  |\vec{q}\, |^3 \frac{\sinh{\vec{a}_2\Cdot \vec{q}}}{\vec{a}_2\Cdot \vec{q}}\\ &= 
 \frac{9}{4\pi} \int_{0}^\infty\!dt \, \left( \frac{1}{[(\vec{b}+ (t-1)\vec{\tilde{a}}_2)^2]^{5/2}} - \frac{1}{[(\vec{b}+ (t+1)\vec{\tilde{a}}_2)^2]^{5/2}} \right) \\
    &= \frac{9}{4\pi} \left( \int_{-1}^\infty\!dt \, \frac{1}{[(\vec{b}+ t\vec{\tilde{a}}_2)^2]^{5/2}} - \int_{1}^\infty dt \frac{1}{[(\vec{b}+ t\vec{\tilde{a}}_2)^2]^{5/2}} \right) \\
    &= \frac{9}{4\pi} \int_{-1}^{+1}\!dt \, \frac{1}{[(\vec{b}+ t\vec{\tilde{a}}_2)^2]^{5/2}} \ ,
    \end{split}
\end{align}
where we have set $\epsilon = 0$ since the integrals are now convergent and we recall that we have defined $\vec{a}_2=i\vec{\tilde{a}}_2$.
The last integral can be done explicitly, giving
\begin{equation}
   \mathcal{J}= \frac{9}{4\pi} \int_{-1}^{1}\!dt \, \frac{1}{(-|\vec{a}_2|^2 t^2 + i 2 \vec{a}_2 \Cdot \vec{b} t + |\vec{b}|^2)^{5/2}} 
   = \frac{3 i \vec{a}_2\Cdot \vec{b}_+ (N-2\vec{a}_{2}^{\, 2} \, \vec{b}_+^{\, 2})}{4\pi N^2 (\vec{b}_+^{\, 2})^{3/2}} + {\rm c.c.}\, , 
\end{equation}
where $\vec{b}_\pm \coloneq \vec{b} \pm i \vec{a_2}$ and $N \coloneq (\vec{a}_2\Cdot \vec{b})^2 - \vec{a}_2^{\, 2} \, \vec{b}^{\, 2} = (1/4)\big[ \vec{b}_+^{\, 2} \vec{b}_-^{\, 2} - (\vec{b}_+\Cdot \vec{b}_-)^2\big]$.

As mentioned earlier, we can now  include the  factor of $\vec{\epsilon}_2 \Cdot\vec{q}$ in \eqref{I2intdef} and \eqref{I2tildeintdef} by writing $\vec{q}$ as a derivative in $\vec{b}$, yielding
\begin{align}
\begin{split}
   \mathcal{E}
   & \coloneq\int\!\frac{d^{D-2}q}{(2\pi)^{D-2}}e^{- i \vec{q}\Cdot \vec{b}}\,  |\vec{q}\, |^3 (\vec{\epsilon}_2 \Cdot \vec{q})  \  \frac{\sinh{\vec{a}_2\Cdot \vec{q}}}{\vec{a}_2\Cdot \vec{q}} = 
   i \, \vec{\epsilon}_2 \Cdot \frac{\partial}{\partial \vec{b}} \,  \mathcal{J}
   \\ & = \frac{3}{4\pi }\frac{\vec{a}_2\Cdot \vec{b}_+ \, \vec{\epsilon}_2 \Cdot \vec{b}}{N (\vec{b}_+^{\, 2})^{3/2}}\left( \frac{3}{\vec{b}_+^2} - \frac{4\vec{a}_2^{\, 2}(N-2\vec{a}_{2}^{\, 2}\,  \vec{b}_+^{\, 2})}{N^2} \right) - {\rm c.c.}\, .
   \end{split}
\end{align}
It is easy to obtain $\mathcal{F}_1$ and $\widetilde{\mathcal{F}}_1$ by applying the shift-operators \eqref{eq:sinhcoshoperators}  on to \eqref{FT-simple}, 
\begin{align}
\mathcal{F}_1 &= \frac{9}{8 \pi}\sum_{\sigma_1=\pm1}
\sum_{\sigma_2=\pm1}\frac{1}{[(\vec{b}+\sigma_1 i\vec{a}_1+\sigma_2 i\vec{a}_2)^2]^{5/2}} \ , \\
\label{Itilde1}
\widetilde{\mathcal{F}}_1 &= \frac{9}{8 \pi}\sum_{\sigma_1=\pm1}
\sum_{\sigma_2=\pm1}\frac{\sigma_1}{[(\vec{b}+\sigma_1 i\vec{a}_1+\sigma_2 i\vec{a}_2)^2]^{5/2}} \ ,
\end{align}
and similarly $\mathcal{F}_2$ and $\widetilde{\mathcal{F}}_2$ are expressible in terms of $\mathcal{J}$ using the shift operators as
\begin{align}
\begin{split}
\mathcal{F}_2 &= \frac{1}{2}\left( \mathcal{E}|_{\vec{b} \to \vec{b}+ i \vec{a}_1} + \mathcal{E}|_{\vec{b} \to \vec{b}- i \vec{a}_1}\right) \ , \\
\widetilde{\mathcal{F}}_2 &= 
\frac{1}{2}\left( \mathcal{E}|_{\vec{b} \to \vec{b}+ i \vec{a}_1} - \mathcal{E}|_{\vec{b} \to \vec{b}- i \vec{a}_1} \right) \ .
\end{split}
\textit{}\end{align}

\subsection{Explicit results and discussion}

We provide the general, somewhat lengthy results for the momentum and spin kicks   in our \href{https://github.com/QMULAmplitudes/Cubic-Corrections-to-Spinning-Observables-from-Amplitudes}{{\it Cubic Corrections to Spinning Observables from Amplitudes} GitHub repository}.   
For illustration, we will now present explicit expressions of these results in a number of special cases. To display the results it is useful to return to four-vector notation for the impact parameter and spin vectors. We also define a projector to the plane orthogonal to $u_1$ and $u_2$ as follows,
\begin{align}
\begin{split}
\label{pimunu}
    \Pi^{\mu\nu}\coloneq \frac{\epsilon(\mu,\rho,u_1,u_2)\epsilon(\nu,\rho,u_1,u_2)}{\sigma^2-1}& =\eta^{\mu\nu}{-}\frac{(u_1^\mu{+}u_2^\mu)(u_1^\nu+u_2^\nu)}{2(\sigma+1)}{+}\frac{(u_1^\mu-u_2^\mu)(u_1^\nu-u_2^\nu)}{2(\sigma-1)}\\ 
    & 
    = \eta^{\mu \nu} -  u_1^\mu \check{u}_1^\nu-  u_2^\mu \check{u}_2^\nu\, , 
    \end{split}
\end{align}
where the dual four-velocities are defined as \cite{Herrmann:2021tct}
\begin{align}
    \check{u}_1^\mu \coloneq \frac{\sigma u_2^\mu - u_1^\mu }{\sigma^2-1}\, , \qquad 
    \check{u}_2^\mu \coloneq \frac{\sigma u_1^\mu - u_2^\mu }{\sigma^2-1}\, , 
\end{align}
with 
\begin{align}
\check{u}_i \Cdot u_j = \delta_{ij}\, , \qquad i,j =1,2\, .
\end{align}
Then the projected impact parameter and spin vectors are
\begin{equation}
\label{baeq}
    b_{\perp}^\mu=\Pi^{\mu\nu}b_{\nu}=(0,\vec{b})^\mu,\qquad a_{i\perp}^\mu=\Pi^{\mu\nu}a_{i\nu}=(0,\vec{a}_i)^\mu\,,
\end{equation}
which allows us to rewrite the results of the previous section,  where the last equalities in both equations above are  specific to   the centre of mass frame. 
\subsubsection{Results for the  momentum kick}
The integrals we need in order to compute the impulse are simply derivatives in $b^\mu$ of the integrals in \eqref{eq: SinchCoshIntegrals2}:
\begin{align}\label{eq: SinchCoshIntegralsImpulse}
\mathcal{B}_1^\mu \coloneq &\int\!\frac{d^{D-2}q_{\perp}}{(2\pi)^{D-2}}e^{ i q_{\perp}\Cdot b_{\perp}} \, q_{\perp}^\mu |q_{\perp}\, |^3 \cosh{a_{1\perp}\Cdot q_{\perp}} \ \cosh{a_{2\perp}\Cdot q_{\perp}} , \\
\tilde{\mathcal{B}}_1^\mu \coloneq &\int\!\frac{d^{D-2}q_{\perp}}{(2\pi)^{D-2}}e^{ i q_{\perp}\Cdot b_{\perp}} \, q_{\perp}^\mu |q_{\perp}\, |^3 \sinh{a_{1\perp}\Cdot q_{\perp}} \ \cosh{a_{2\perp}\Cdot q_{\perp}} , \\
\label{B2intdef}
\mathcal{B}_2^\mu \coloneq    &\int\!\frac{d^{D-2}q_{\perp}}{(2\pi)^{D-2}}e^{ i q_{\perp}\Cdot b_{\perp}}\, q_{\perp}^\mu  |q_{\perp}\, |^3 (\epsilon_2 \Cdot q_{\perp}) \cosh{a_{1\perp}\Cdot q_{\perp}} \  \frac{\sinh{a_{2\perp}\Cdot q_{\perp}}}{a_{2\perp}\Cdot q_{\perp}} , 
\\
\label{B2tildeintdef}
\tilde{\mathcal{B}}_2^\mu \coloneq    &\int\!\frac{d^{D-2}q_{\perp}}{(2\pi)^{D-2}}e^{ i q_{\perp}\Cdot b_{\perp}}\, q_{\perp}^\mu |q_{\perp}\, |^3 (\epsilon_2 \Cdot q_{\perp}) \sinh{a_{1\perp}\Cdot q_{\perp}} \ \frac{\sinh{a_{2\perp}\Cdot q_{\perp}}}{a_{2\perp}\Cdot q_{\perp}} \, , 
\end{align}
where $\epsilon_2 \Cdot q_{\perp} \coloneq \epsilon(u_1,u_2,a_2,q)$. The results for these integrals can be written using the expressions 
\begin{align}
    b_{\sigma_1\sigma_2}^\mu\coloneq b_\perp^\mu+\sigma_1 ia_{1\perp}^\mu+\sigma_2 ia_{2\perp}^\mu\,,\quad N_{\sigma_1}\coloneq (a_{2\perp}\Cdot b_{\sigma_1 0})^2-a_{2\perp}^2 b_{\sigma_10}^2\, ,
\end{align}
and
\begin{align}
\begin{split}
        &c_{\sigma_1\sigma_2}^\mu\coloneq  \frac{1}{  
N_{\sigma_1}^4 } \Big[ 
\epsilon (\mu a_{2\perp} u_1 u_2)N_{\sigma_1} b_{{\sigma_1\sigma_2}}^2 a_{2\perp}\Cdot 
b_{{\sigma_1\sigma_2}}\big(3 
N_{\sigma_1}^2{-}4 a_{2\perp}^2 b_{{\sigma_1\sigma_2}}^2 
\big(N_{\sigma_1}{-}2 a_{2\perp}^2 
b_{{\sigma_1\sigma_2}}^2\big)\big) \\
&\quad{+}3\epsilon (a_{2\perp} b_{\perp} u_1 u_2)\Big(b_{\sigma_1 \sigma_2}^{\mu } a_{2\perp}\Cdot 
b_{{\sigma_1\sigma_2}} 
 \big(5 N_{\sigma_1}^3{-}2 a_{2\perp}^2 
b_{{\sigma_1\sigma_2}}^2 \big(4 a_{2\perp}^2 b_{{\sigma_1\sigma_2}}^2 
\big(2 a_{2\perp}^2 b_{{\sigma_1\sigma_2}}^2{-}N_{\sigma_1}\big){+}3 
N_{\sigma_1}^2\big)\big) \\
&\quad{+}a_{2\perp}^{\mu } b_{{\sigma_1\sigma_2}}^2 
\big(2 a_{2\perp}^2 b_{{\sigma_1\sigma_2}}^2 \big(4 a_{2\perp}^2 
b_{{\sigma_1\sigma_2}}^2 \big(2 a_{2\perp}^2 
b_{{\sigma_1\sigma_2}}^2{+}{N_{\sigma_1}}\big){-}N_{\sigma_1}^2\big){+}N_{\sigma_1}^3\big)\Big)\Big]\, .
\end{split}
\end{align}
Explicitly we have
\begin{align}
&\mathcal{B}_1^\mu=  -\frac{45 i}{8 \pi}\sum_{\sigma_1=\pm1}
\sum_{\sigma_2=\pm1}\frac{b_{\sigma_1\sigma_2}^\mu}{|b_{\sigma_1\sigma_2}|^{7}} \ ,  \\
&\tilde{\mathcal{B}}_1^\mu=   -\frac{45i}{8 \pi}\sum_{\sigma_1=\pm1}
\sum_{\sigma_2=\pm1}\frac{ (-\sigma_1)\,b_{\sigma_1\sigma_2}^\mu}{|b_{\sigma_1\sigma_2}|^{7}} \ ,\\
&\mathcal{B}_2^\mu = \frac{3 i}{8 \pi  }
\sum_{\sigma_1=\pm1}
\sum_{\sigma_2=\pm1}
\frac{\sigma_2 c^\mu_{\sigma_1\sigma_2}}{ |b_{\sigma_1\sigma_2}|^{7}}\, , 
\\
&\tilde{\mathcal{B}}_2^\mu=\frac{3 i}{8 \pi }
\sum_{\sigma_1=\pm1}
\sum_{\sigma_2=\pm1}
\frac{(-\sigma_1)\sigma_2 c^\mu_{\sigma_1\sigma_2}}{ |b_{\sigma_1\sigma_2}|^{7}}\, , 
\end{align}
which are the building blocks of the impulse
\begin{align}
    \langle  \Delta \mathbb{P}_1^\mu  \rangle_{I_1} &=i \left(\frac{\kappa}{2}\right)^6\frac{3}{64}\frac{m_1^2 m_2}{\sqrt{\sigma^2-1}}\Big((\sigma^2-1)\mathcal{B}^\mu_1-i\sigma\mathcal{B}^\mu_2\Big)+(1\leftrightarrow2)\,,\\
    \langle  \Delta \mathbb{P}_1^\mu  \rangle_{G_3} &=-i \left(\frac{\kappa}{2}\right)^6\frac{3}{64}\frac{m_1^2 m_2}{\sqrt{\sigma^2-1}}\mathcal{B}^\mu_1+(1\leftrightarrow2)\,,\\
    \langle  \Delta \mathbb{P}_1^\mu  \rangle_{\tilde{I}_1} &=- \left(\frac{\kappa}{2}\right)^6\frac{3}{64}\frac{m_1^2 m_2}{\sqrt{\sigma^2-1}}\Big((\sigma^2-1)\tilde{\mathcal{B}}^\mu_1-i\sigma\tilde{\mathcal{B}}^\mu_2\Big)-(1\leftrightarrow2)\,,\\
    \langle  \Delta \mathbb{P}_1^\mu  \rangle_{\tilde{G}_3} &= \left(\frac{\kappa}{2}\right)^6\frac{3}{64}\frac{m_1^2 m_2}{\sqrt{\sigma^2-1}}\tilde{\mathcal{B}}^\mu_1-(1\leftrightarrow2)\,.
\end{align}
By $(1\leftrightarrow2)$ here we mean swapping $a_1\leftrightarrow a_2$ and $m_1\leftrightarrow m_2$. 
The exact results for the spin kick are rather complicated, hence we will only present expressions that are either expanded to linear order in the spins in Section~\ref{sec:expinspin}, or in the aligned spin case in Section~\ref{sec:aligned}. 

\subsubsection{Expanded in spin momentum  and spin kicks} 
\label{sec:expinspin}
The leading-order momentum kick of particle $1$ due to the $R^3$ corrections can be obtained from  the general expression \eqref{dp1}.  Expanding to linear order in the spins we get, for the parity-even and parity-odd cubic deformations:
\begin{align}
\label{imp1}
    \langle  \Delta \mathbb{P}_1^\mu  \rangle_{I_1} &= \left(\frac{\kappa}{2}\right)^6\frac{135}{128\pi} \bigg\{ 
    \sqrt{\sigma^2-1}  m_1 m_2 (m_1 + m_2) \frac{b_\perp^\mu}{| b_\perp|^7} \notag\\
    &- 
    \frac{\sigma }{\sqrt{\sigma^2-1}} \frac{m_1 m_2}{|  b_\perp|^9}\Big[  7b_\perp^\mu \big( m_1\epsilon(u_1 u_2 a_2 b_\perp)+m_2\epsilon(u_1 u_2 a_1 b_\perp)\big)\notag\\
    &+|b_\perp|^2 \big(m_1\epsilon(u_1 u_2 a_2 \mu)+m_2\epsilon(u_1 u_2 a_1 \mu)\big)\Big] \bigg\} \, ,\\
    \nn\\
    \langle  \Delta \mathbb{P}_1^\mu  \rangle_{G_3} &= -\left(\frac{\kappa}{2}\right)^6\frac{135}{128\pi} \frac{m_1 m_2 (m_1 + m_2)}{\sqrt{\sigma^2-1}}   \frac{b_\perp^\mu}{|  b_\perp|^7} \,,\\
    \nn\\
    \langle  \Delta \mathbb{P}_1^\mu  \rangle_{\tilde{I}_1} &= \left(\frac{\kappa}{2}\right)^6\frac{135}{128\pi} \sqrt{\sigma^2-1}\frac{m_1 m_2}{|b_\perp|^9}\Big[7b^\mu_\perp(m_1 a_1\Cdot b_\perp - m_2 a_2\Cdot b_\perp)\nn\\
    &+|b_\perp|^2(m_1 a_{1\perp}^\mu- m_2 a_{2\perp}^\mu)\Big] \nn\\
    &=\left(\frac{\kappa}{2}\right)^6\frac{135}{128\pi} \sqrt{\sigma^2-1}\frac{m_1 m_2}{|b_\perp|^9}\big(S_{1\nu}-S_{2\nu}\big)\big(|b_\perp|^2\Pi^{\mu\nu}+7b_{\perp}^\mu b_{\perp}^\nu \big)\, , \\
    \nn\\
    \label{imp4}
    \langle  \Delta \mathbb{P}_1^\mu  \rangle_{\tilde{G}_3} &= -\left(\frac{\kappa}{2}\right)^6\frac{135}{128\pi} \frac{1}{\sqrt{\sigma^2-1}}\frac{m_1 m_2}{|b_\perp|^9}\Big[7b^\mu_\perp(m_1 a_1\Cdot b_\perp - m_2 a_2\Cdot b_\perp)\nn\\
    &+|b_\perp|^2(m_1 a_{1\perp}^\mu- m_2 a_{2\perp}^\mu)\Big]  \nn\\
    &=-\left(\frac{\kappa}{2}\right)^6\frac{135}{128\pi} \frac{1}{\sqrt{\sigma^2-1}}\frac{m_1 m_2}{|b_\perp|^9}\big(S_{1\nu}-S_{2\nu}\big)\big(|b_{\perp}|^2\Pi^{\mu\nu}+7b_{\perp}^\mu b_{\perp}^\nu \big)\, , 
\end{align}
where $|b_\perp|=\sqrt{-b_\perp\Cdot b_\perp}$, and $\Pi^{\mu \nu}$ was defined in \eqref{pimunu}. Note that the corrections due to spin for $G_3$ occur only at quadratic order in the $a_i$.

Using \eqref{spingchange1} we get, for the spin kick of particle $1$ expanded to first order in the spins $a_i$, 
\begin{align}
\label{sk1}
    \langle  \Delta \mathbb{S}_1^\mu  \rangle_{I_1} &=
    \left(\frac{\kappa}{2}\right)^6\frac{135}{128\pi}\frac{1}{\sqrt{\sigma^2-1}}\frac{m_1 m_2}{|b_\perp|^{7}} \Big[m_2\sigma\, a_1\Cdot u_2  b_\perp^\mu \notag
    \\ & +a_{1\perp}\Cdot b_\perp\big[- (\sigma^2-1)m_1 u_1^\mu +m_2 (u_1^\mu -\sigma u_2^\mu)\big]\Big] \, ,\\
    \label{sk2}
    \langle  \Delta \mathbb{S}_1^\mu  \rangle_{G_3} 
  &=
    \left(\frac{\kappa}{2}\right)^6\frac{135}{128\pi}\frac{1}{\sqrt{\sigma^2-1}}\frac{m_1 m_2}{|b_\perp|^{7}}(m_1+m_2)(a_{1\perp}\Cdot b_\perp)u_1^\mu\, , 
    \\ 
    \label{sk3}
     \langle  \Delta \mathbb{S}_1^\mu  \rangle_{\tilde{I}_1} &=
     \left(\frac{\kappa}{2}\right)^6\frac{135}{128\pi}\sqrt{\sigma^2-1}\frac{m_1^2 m_2}{|b_\perp|^{7}} \epsilon(u_1a_{1}b_\perp\mu)\, ,  \\
     \label{sk4}
     \langle  \Delta \mathbb{S}_1^\mu  \rangle_{\tilde{G}_3} &=
     -\left(\frac{\kappa}{2}\right)^6\frac{135}{128\pi}\frac{1}{\sqrt{\sigma^2-1}}\frac{m_1^2 m_2}{|b_\perp|^{7}} \epsilon(u_1a_{1}b_\perp\mu)\, .
\end{align} 
Note that the expressions of the impulse \eqref{imp1}-\eqref{imp4} only depend on the spin via the projections  $a_{i\perp}$ defined in \eqref{baeq}, satisfying $u_i\Cdot a_{j \perp} {=}0$. This also happens for the  spin kick \eqref{sk2} for the $G_3$ deformation at linear order in the spin, however at $\cO(a_i^2)$ and beyond this is no longer true. The spin kicks associated to $I_1$, $\tilde{I}_1$ and $\tilde{G_3}$ in \eqref{sk1}, \eqref{sk3} and  \eqref{sk4} generically depend on all the components of the four-vectors $a_i^\mu$.

\subsubsection{Aligned spin case}
\label{sec:aligned}
An interesting situation to discuss is the aligned-spin case, where 
\begin{align}
p_i \Cdot a_j = 0\, , \qquad {a}_i\Cdot {b_\perp} =0\, , \qquad 
i,j=1,2\, .  
\end{align}
For instance, as in \cite{Vines:2017hyw}
one could pick  an orthonormal basis 
$\{e^\mu_{\hat{a}}\}_{a=0}^3$, defined such that  $\eta_{\mu \nu} e^\mu_{\hat{a}} e^\nu_{\hat{b}} {=} \delta_{ab}$ and $\epsilon_{\mu \nu \alpha \beta}e^\mu_{\hat{0}}e^\nu_{\hat{1}}e^\alpha_{\hat{2}}e^\beta_{\hat{3}}=1$, and align the spins as $a^\mu_{1,2} {=} a_{1,2} \, e^\mu_{\hat{3}}$, the  impact parameter as $b^\mu {=} b \, e^\mu_{\hat{1}}$, 
with the velocities $u_1$ and $u_2$ being in the $(\hat{0}, \hat{2})$ plane.
In this configuration  the spin vectors of the two bodies are then aligned with the orbital (and hence also total) angular momentum of the system. Note that in the aligned case  the four-vectors  $a_{i}$ and $a_{i \perp}$ actually coincide.

The  aligned-spin expressions for $\langle  \Delta \mathbb{P}_1^\mu  \rangle_{I}$ and 
$\langle  \Delta \mathbb{S}_1^\mu  \rangle_{I}$, for $I\in (I_1,  G_3, \tilde{I}_1, \tilde{G}_3)$, are found to be  schematically of the form: 
\begin{align}
\label{impuls-aligned}
   \langle \Delta \mathbb{P}_1^\mu  \rangle_{I_1, G_3} \propto b^\mu_\perp\, , \qquad 
   \langle \Delta \mathbb{P}_1^\mu  \rangle_{\tilde{I}_1, \tilde{G}_3}\propto a^\mu\, , 
\end{align}
where $a^\mu$ is the common direction of $a_{1,2}^\mu$, and 
\begin{align}
   \langle \Delta \mathbb{S}_1^\mu  \rangle_{I_1, G_3} =0 \, , \qquad  \ 
   \langle \Delta \mathbb{S}_1^\mu  \rangle_{\tilde{I}_1, \tilde{G}_3}= C_{\tilde{I}_1, \tilde{G}_3} u_1^\mu +D_{\tilde{I}_1, \tilde{G}_3}\epsilon(u_1a_{1}b\mu)\, , 
\end{align}
where $C_i,\,D_i$ are complicated kinematic-dependent factors. 

A few comments are in order here. First, we note that for parity-even interactions the motion lies on the plane defined by the relative momentum $\vec{p}$ and the impact parameter $\vec{b}_\perp$, and the spin is conserved.
On the other hand, for parity-odd interactions the situation is substantially different: the  impulse of particle $1$ lies in the direction of the spins, hence the motion is nonplanar,  see also 
\cite{Cano:2019ore,Datta:2020axm,Fransen:2022jtw} for related observations.

If the orbits lie on a plane, one can also define a scattering angle $\chi$,  given by 
(see e.g.~\cite{Guevara:2018wpp,Parra-Martinez:2020dzs})
\begin{align}
\label{eq:scatteringangle}
    \sin \frac{\chi}{2} = - \frac{1}{2 {| \vec{p}\, |}}\frac{\partial}{\partial b } \delta_{\text{as}}\, , 
\end{align}
where
\begin{align}
|\vec{p}\, | = \frac{m_1 m_2 \sqrt{\sigma^2 -1}}{E} = \frac{m_1 m_2 \sqrt{\sigma^2 -1}}{\sqrt{m_1^2 + m_2 ^2 + 2 m_1 m_2 \sigma}}\, 
\end{align}
is the (common) momentum in the centre of mass frame, and $\delta_{\text{as}} \coloneqq \delta \big|_{p_i\cdot a_j = a_i\cdot b_\perp = 0}$ with $\delta(b) = - i \cM_4(b)$, where $\cM_4(b)$ is the one-loop elastic amplitude in impact-parameter space. Furthermore, 
since we are working to leading order in the perturbations,  we can approximate $\sin (\chi/2) \approx \chi/2$. We denote $\chi_{\mathcal{F}_1}$ and $\chi_{\mathcal{F}_2}$ the contributions to the scattering angle from the Fourier transforms $\mathcal{F}_1$ and $\mathcal{F}_2$ in \eqref{eq: SinchCoshIntegrals2} respectively, and obtain 
\begin{align}
    \chi_{\mathcal{F}_1} &= \frac{45 |b| E}{4\pi \sqrt{\sigma^2-1}m_1 m_2} \Bigg\{ \frac{1}{(b^2-a_+^2)^{7/2}} + \frac{1}{(b^2-a_-^2)^{7/2}} \Bigg\}\\
    \chi_{\mathcal{F}_2} &= i\frac{15 E}{4 \pi m_1 m_2}\Bigg\{ \frac{a_+ (18 b^6-35a_+^2 b^4+28a_+^4 b^2-8a_+^6 )}{b^6(b^2-a_+^2)^{7/2}} \nn \\
    &-\frac{a_- (18 b^6-35a_-^2 b^4+28a_-^4 b^2-8a_-^6 )}{b^6(b^2-a_-^2)^{7/2}} \Bigg\}\, .
\end{align}
   where $a_+=a_1+a_2$ and $a_-=a_1-a_2$. The deflection angle for the $I_1$ and $G_3$ interactions can be written in terms of these building blocks:
   \begin{align}
       \chi_{I_1} =& \left(\frac{\kappa}{2}\right)^6 \frac{3}{64} \frac{m_1^2 m_2}{ \sqrt{\sigma^2-1}} \left[(\sigma^2-1) \chi_{\cF_1} - i \sigma \chi_{\cF_2}\right] +(1\leftrightarrow 2)\,,\\
       \chi_{G_3} =&  \left(\frac{\kappa}{2}\right)^6 \frac{3}{64} \frac{m_1^2 m_2}{ \sqrt{\sigma^2-1}} (-\chi_{\cF_1})+(1\leftrightarrow 2)\, ,
   \end{align}
where by $(1\leftrightarrow2)$ we mean swapping $m_1{\leftrightarrow} m_2$, $a_{+}{\rightarrow} a_{+}$ and $a_{-}{\rightarrow}-a_{-}$.
For the parity-odd case, applying \eqref{eq:scatteringangle} yields
\begin{align}
    \chi_{\tilde{\mathcal{F}}_1} =& \chi_{\tilde{\mathcal{F}}_2} = 0 \, .  
\end{align}
This reflects the fact that the particles are deflected in the perpendicular direction to the scattering plane, as we can see from \eqref{impuls-aligned}.

\section{Conclusions}
In this paper we have studied the effect of all independent parity-even and parity-odd cubic deformations on the dynamics of two Kerr black holes. 
When looking at two-to-two scattering, the leading-order corrections   first appear at one loop.  By including the effects of spin, we obtained the relevant Compton amplitudes, an essential ingredient for the calculation -- which we performed both in the leading singularity and HEFT approaches. With the elastic amplitude in hand, we computed a number of  classical observables using the KMOC formalism:  the impulse, the spin kick, and the scattering angle for all cubic deformations; their full expressions can be found in the \href{https://github.com/QMULAmplitudes/Cubic-Corrections-to-Spinning-Observables-from-Amplitudes}{{\it Cubic Corrections to Spinning Observables from Amplitudes} GitHub repository}.
Examining these in the aligned spin limit, we observe planar motion for the parity-even interactions, as in the  Einstein-Hilbert case. On the other hand, parity-odd interactions lead to non-planar scattering, kicking the black holes off the 
scattering plane.

There are several interesting directions to examine further. A natural next step is to look at how these cubic deformations affect the waveform when spin is included. In particular, the parity-odd terms are of interest due to their  peculiar effects on the dynamics of the spinning objects.  As mentioned in the introduction, in the scalar case the parity-even interaction $G_3$ can can be reinterpreted as a tidal
deformation after a field redefinition, as hinted at by the lack of a pole in $\eqref{G3amplitude}$. It would be interesting to see if this statement can be extended to the spinning case for both $G_3$ and its parity-odd version $\tilde{G}_3$, with the latter  not contributing to the four-point scalar amplitude at one loop. Another avenue is looking at how other higher-derivative corrections, such as $R^4$ interactions, behave in the presence of  spin. 
We will come back to these questions in the near future.

\section*{Acknowledgements}
 We would like to thank  
 Lara Battino, Lara Bohnenblust, 
 Lucile Cangemi, Gang Chen, Joshua Gowdy, Henrik Johansson,  Scott Melville and  Nathan Moynihan for  useful conversations. 
This work was supported by the Science and Technology Facilities Council (STFC) Consolidated Grants   ST/T000686/1 and  ST/X00063X/1 \textit{``Amplitudes, Strings  \& Duality''}.
The work of GRB and PVM is supported by STFC quota studentships.   GT is also supported by a Leverhulme research fellowship RF-2023-279$\backslash 9$.
No new data were generated or analysed during this study.

\newpage

\bibliographystyle{JHEP}
\bibliography{ScatEq.bib}
\end{document}